\documentclass[aps,11pt,preprint,showpacs,preprintnumbers,nofootinbib]{revtex4-1}
\usepackage{graphicx}
\usepackage{dcolumn}
\usepackage{bm,amsfonts,amsthm,amsmath,amssymb}
\usepackage[]{epsfig,graphics}
\usepackage{comment}
\usepackage[T1]{fontenc}
\usepackage{lipsum}
\usepackage{ulem}
\usepackage{multirow}
\usepackage{color}
\usepackage{lastpage}
\usepackage{enumerate}
\usepackage{subfigure}
\usepackage{wasysym}
\usepackage{graphicx}
\usepackage[font=small,it,justification=raggedright,singlelinecheck=false]{caption}

\usepackage{hyperref}

\hypersetup{
    bookmarks=true,         
    unicode=false,          
    pdftoolbar=true,        
    pdfmenubar=true,        
    pdffitwindow=false,     
    pdfstartview={FitH},    
    pdftitle={My title},    
    pdfauthor={Author},     
    pdfsubject={Subject},   
    pdfcreator={Creator},   
    pdfproducer={Producer}, 
    pdfkeywords={keyword1} {key2} {key3}, 
    pdfnewwindow=true,      
    colorlinks=false,       
    linkcolor=red,          
    citecolor=green,        
    filecolor=magenta,      
    urlcolor=cyan           
}

\newenvironment{itemize*}
  {\begin{itemize}
    \setlength{\itemsep}{0pt}
    \setlength{\parskip}{0pt}}
  {\end{itemize}}

\newenvironment{enumerate*}
  {\begin{enumerate}
    \setlength{\itemsep}{0pt}
    \setlength{\parskip}{0pt}}
  {\end{enumerate}}

\newenvironment{description*}
  {\begin{description}
    \setlength{\itemsep}{0pt}
    \setlength{\parskip}{0pt}}
  {\end{description}}

\def\d{{\rm d}}

\def\eg{{\it e.g.~}}

\def\ie{{\it i.e.~}}
\def\ben{\begin{enumerate*}}
\def\een{\end{enumerate*}}
\def\bi{\begin{itemize*}}
\def\ei{\end{itemize*}}
\def\bd{\begin{description*}}
\def\ed{\end{description*}}
\def\be{\begin{equation}}
\def\ee{\end{equation}}
\def\bea{\begin{eqnarray}}
\def\eea{\end{eqnarray}}
\def\bfl{\begin{flushleft}}
\def\efl{\end{flushleft}}

\newcommand{\frw}{{\mbox{\tiny FRW}}}
\newcommand{\osc}{{\mbox{\tiny osc}}}

\newcommand{\gsim}{\lower.7ex\hbox{$\;\stackrel{\textstyle>}{\sim}\;$}}
\newcommand{\lsim}{\lower.7ex\hbox{$\;\stackrel{\textstyle<}{\sim}\;$}}
\newcommand{\sgm}{\sigma}
\newcommand{\beq}{\begin{equation}}
\newcommand{\eeq}{\end{equation}}
\newcommand{\fr}{\frac}

\newcommand{\nn}{\nonumber}

\begin{document}

\title{Toward an Effective Field Theory Approach to Reheating}

\author{Ogan \"{O}zsoy${}^{1}$}
\email{oozsoy@syr.edu} 
\author{John T. Giblin, Jr.${}^{2,3}$}
\email{giblinj@kenyon.edu}
\author{Eva Nesbit${}^{1}$}
\email{ehnesbit@syr.edu}
\author{Gizem \c{S}eng\"{o}r${}^{1}$}
\email{gsengor@syr.edu} 
\author{Scott Watson${}^{1,4}$}
\email{gswatson@syr.edu} 

\affiliation{${}^1$Department of Physics, Syracuse University, Syracuse, NY 13244}
\affiliation{${}^2$Department of Physics, Kenyon College, 201 N College Rd, Gambier, OH 43022}
\affiliation{${}^3$CERCA/ISO, Department of Physics, Case Western Reserve University, 10900 Euclid Avenue, Cleveland, OH 44106}
\affiliation{${}^4$Department of Physics, Tokyo Institute of Technology, Tokyo 152-8551, Japan}

\date{\today}

\begin{abstract}
We investigate whether
Effective Field Theory (EFT) approaches, which have been useful in examining inflation and dark energy, can also be used to establish a systematic approach to inflationary reheating.
We consider two methods. First, we extend Weinberg's background EFT to the end of inflation and reheating.
We establish when parametric resonance and decay of the inflaton occurs, but also find intrinsic theoretical limitations, which make it difficult to capture some reheating models.
This motivates us to next consider Cheung, {\it et. al.}'s EFT approach, which instead focuses on perturbations and the symmetry breaking induced by the cosmological background.  Adapting the latter approach to reheating implies some new and important differences compared to the EFT of Inflation.
In particular, there are new hierarchical scales, and we must account for inflaton oscillations during reheating, which lead to discrete symmetry breaking.
Guided by the fundamental symmetries, we construct the EFT of reheating, and as an example of its usefulness we establish a new class of reheating models and the corresponding predictions for gravity wave observations.
In this paper we primarily focus on the first stages of preheating. We conclude by discussing challenges for the approach and future directions. 
This paper builds on ideas first proposed in the note \cite{Ozsoy:2015rna}.
\end{abstract}
\maketitle
\thispagestyle{empty}
\tableofcontents
\section{Introduction}
If inflation occurred in the early universe it must have eventually ended resulting in a hot, thermal universe by the time of Big Bang Nucleosynthesis (BBN).
The process by which the inflaton's energy is transferred into other particles -- which hopefully, eventually, gave rise to Standard Model particles -- is known as inflationary reheating.
Reheating can occur perturbatively \cite{Abbott:1982hn,Albrecht:1982mp,Dolgov:1982th}, or non-perturbatively in a process known as preheating \cite{Dolgov:1989us,Traschen:1990sw,Kofman:1997yn} (see \cite{Amin:2014eta,Allahverdi:2010xz} for recent reviews). 

Existing investigations into reheating have been rather model dependent, often focusing on constraining the precise regions of the parameter space that lead to successful reheating.
Analytic methods for exploring the dynamics still rely on the earliest works mentioned above, and the non-linearities and complexity of the reheating process still require invoking numeric/lattice methods \cite{Khlebnikov:1996mc,Easther:2006gt,GarciaBellido:1997wm,Felder:2000hj,Easther:2006vd,Price:2008hq,Easther:2010qz,Amin:2014eta,Allahverdi:2010xz}. 
Moreover, the wealth of cosmological observations from the Cosmic Microwave Background (CMB) and Large Scale Structure (LSS) relate to the physics of inflation far before reheating, and so the lack of observational windows on (p)reheating has also made its study far less compelling than inflation -- with the prediction of gravitational waves providing a possible exception. 

In this paper, we take steps to address the model dependence of (p)reheating building on motivation from recent works \cite{Hertzberg:2014iza,Amin:2015ftc,Ozsoy:2015rna}.  Our approach is to use the Effective Field Theory (EFT) approach to cosmology, which at this point has been applied to all cosmic epochs except for (p)reheating.  We will first consider the EFT of the background as developed by Weinberg for inflation in \cite{Weinberg:2008hq} and later adapted to studies of dark energy in \cite{Park:2010cw}.  Ultimately, we will find that this approach is not completely satisfactory in generalizing studies of reheating.  Instead we find that the different approach of the EFT of cosmological perturbations is more promising. 

The EFT of Inflation \cite{Creminelli:2006xe,Cheung:2007st,Senatore:2010wk} and generalizations to dark energy \cite{Creminelli:2008wc,Bloomfield:2012ff,Gubitosi:2012hu,Fasiello:2016qpn,Fasiello:2016yvr,Lewandowski:2016yce} and structure formation \cite{Baumann:2010tm} are based on the idea that there is a physical clock corresponding to the Goldstone boson that non-linearly realizes the spontaneously broken time diffeomorphism
invariance of the background.  In unitary gauge -- where the clock is homogeneous -- the matter perturbations are encoded within the metric, \ie the would-be Goldstone boson is `eaten' by the metric,  since  gravity is a gauge theory.   After we establish the limitations of the EFT background approach, we then present an EFT of reheating using this EFT of perturbations to develop a more robust approach to studying the end of inflation and reheating.

The rest of the paper is as follows. In Section \ref{sec1}, we review some of the important issues and constraints surrounding particular examples of (p)reheating models.  In Section \ref{sec2}, we consider Weinberg's approach to the EFT of Inflation, and consider how inflation might end and (p)reheating would proceed.  We find that the perturbative approach to the background presents a substantial challenge to this approach, along with the usual problem of knowing the complete inflationary potential.  This motivates us to construct an EFT of reheating in Section \ref{sec3} -- focusing on the EFT of the perturbations.  We analyze the process of particle production, demonstrate how our approach connects to existing preheating models, and discuss ways in which our EFT can be used to connect to both inflation (and its end) and observations.  In Section \ref{sec4}, we conclude and discuss the challenges facing our approach and future directions.

\section{Challenges for Inflationary Reheating \label{sec1}}
Model dependent studies of (p)reheating have raised a number of important questions and issues. 
From the perspective of inflationary model building within string theory, the requirement to isolate the inflationary sector to achieve an adequate duration of inflation can result in challenges in transferring the energy density to other fields, and eventually the Standard model sector following inflation \cite{Kofman:2005yz}.  The complexity of the string landscape and the large number of moduli fields can exacerbate this problem \cite{Green:2007gs}.
In bottom-up approaches, toy models often demonstrate a conflict between the need for the inflaton to have feeble interactions during inflation (so as to be consistent with both successful inflation and constraints on non-Gaussianity), and later having strong enough couplings for the complete decay of the inflaton and the (eventual) successful reheating of the Standard Model.  Perturbative decay can also present a challenge depending on the effective mass of the decay channels and the time dependence of the inflaton decay rate \cite{Felder:1998vq}.

As an example, consider Chaotic inflation with $V \sim m^2_\phi \phi^2$ and reheating with a 
renormalizable coupling to a reheat field, $\chi$.  We note that this model is in tension with existing CMB constraints, but it presents a simple example of the more general problems one might anticipate with (p)reheating.
The Lagrangian we consider is\footnote{We work in reduced Planck units $m_{\rm pl}=1/\sqrt{8 \pi G}=2.4 \times 10^{18}\,{\rm GeV}$ with $\hbar=c=1$ and with a `mostly plus' $(-,+,+,+)$ sign convention for the metric. Our conventions for curvature tensors are those of Weinberg.}
\be\label{toyL}
{\cal L}= - \frac{1}{2} \left( \partial \phi \right)^2  - \frac{1}{2} m_\phi^2 \phi^2 -\frac{1}{2} \left( \partial \chi \right)^2-U(\chi) - \fr{g^2}{2} \phi^2\chi^2 ,
\ee
where we assume that initially the reheat field is fixed by its $U(\chi)$ and remains in its vacuum during inflation.
The mass of the inflaton is fixed by the power spectrum \cite{Ade:2015lrj},
\beq
\Delta_{\mathcal{R}}^2 = \fr{1}{96\pi^2}\left(\fr{m_\phi}{m_{\rm pl}}\right)^2 \left(4 N_*\right)^2 \equiv 2.2 \times 10^{-9}
\eeq
where $N_*$ is the number of e-folds before the end of inflation and with $N_* = 60$ we have $m_\phi \simeq 6.4 \times 10^{-6} \, m_{\rm pl}$.
The inflaton will begin to oscillate around the minimum of its potential when its mass becomes comparable to the Hubble scale, $m_\phi \approx H(t_{\rm \osc})$,  with a profile given by the expression $\phi_{0}(t) = \Phi(t) \sin(m_\phi t)$ \cite{Kofman:1997yn}. The amplitude of the oscillations, $\Phi(t)$, is a monotonic function of cosmic time given by $\Phi = \sqrt{8/3}~ (m_{\rm pl}/2\pi N_{\rm \osc})$, where $N_{\rm \osc}$ is the number of oscillations after the end of inflation. Setting $N_{\rm \osc}=1$ gives $\Phi \approx 0.3 \, m_{\rm pl}$, which we take as the initial amplitude of the inflaton oscillations. 

If the direct coupling in \eqref{toyL} presents the only decay channel for the inflaton the expansion of the universe will prevent the complete perturbative decay of the inflaton \cite{Kofman:1997yn}. This is because the decay rate, $\Gamma$, scales as $\Gamma\propto \Phi^2 \sim 1/t^2$ whereas the expansion rate during reheating scales as $H\sim 1/t$.
Instead, in this case decay must proceed non-perturbatively through preheating \cite{Dolgov:1989us,Traschen:1990sw,Kofman:1997yn}, where parametric resonance can lead to enhanced decay of the inflaton condensate.
The mode equation for $\chi$ fluctuations resulting from \eqref{toyL} in the presence of the oscillating condensate $\phi_{_0}(t)$ is
\be \label{mode_eqn}
\ddot{\chi}_k + \left[ k^2 +m_\chi^2 +g^2 \phi_{_0}^2 \right] \chi_k=0,
\ee
where we have neglected the expansion of the universe ($a=1$) and note that including gravitational effects would act to strengthen the main conclusion below.
If the field begins in its Bunch-Davies vacuum the corresponding WKB solution is $\chi_k \sim \exp(-i \int \omega_k(t^\prime) dt^\prime)$, where $\omega_k$ is time-dependent frequency corresponding
to the terms inside the brackets in \eqref{mode_eqn}.
Particle production occurs if the adiabatic conditions 
fail corresponding to $\dot{\omega}_k \gg \omega_k^2$ or $\ddot{\omega}_k \gg \omega_k^3$, etc...
Thus, a necessary condition for preheating is
\be \label{adiabatic}
\frac{\dot{\omega}_k}{\omega_k^2} \simeq \frac{g^2 \phi \dot{\phi}}{\left( k^2 + m_\chi^2 +g^2 \phi^2 \right)^{3/2}} >1,
\ee
corresponding to the production of modes with their momenta satisfying
\be \label{kmodes}
k^2 \lesssim \left( g^2 \phi \dot{\phi}\right)^{2/3}- g^2 \phi^2 -m_\chi^2.
\ee
The ratio in \eqref{adiabatic} is maximal when the inflaton is near the bottom of the potential, where we can approximate $\dot{\phi}_0 \simeq m_\phi \Phi$.
Broad resonance \cite{Kofman:1997yn} will assure us that preheating is successful.  This corresponds to a restriction on the range of wave numbers in the resonance band $\Delta k \gg m_\phi$ Maximizing the right side of \eqref{kmodes} with respect to $\phi$, we find the maximum value of $\phi_*^2 \simeq 0.2 \, \dot{\phi} / g$ corresponding to a maximum value of resonant  momentum $k^2_\ast = 0.4 \, g \dot{\phi}-m_\chi^2$.  Therefore the condition for broad resonance $\Delta k \simeq k_\ast \gg m_\phi$ can be written as a condition on the coupling constant $g$,
\be
 g  \gg  \frac{m_\phi^2 +m_\chi^2}{  \dot{\phi}} \simeq   \frac{m_\phi^2 +m_\chi^2}{  m_\phi \Phi}.
\ee
Taking $\Phi \simeq 0.3 \, m_{\rm pl}$ and assuming $m_\chi \ll m_\phi$
we find $g \gg 3.8 \times 10^{-5}$ for efficient preheating in the broad resonance regime.

On the other hand, we can obtain a lower bound on the strength of the coupling by requiring the one-loop correction induced by the $g^2 \phi^2 \chi^2$ interaction to not to spoil the flatness of the potential during inflation.  That is, we require $\delta m_\phi \lesssim m_\phi \simeq 6.4 \times 10^{-6} \, m_{\rm pl}$, whereas the loop correction is $\delta m^2_\phi = (g^2 \Lambda_{\rm uv}^2) / (16 \pi^2)$. The cut-off is expected to be Planckian $\Lambda_{\rm uv} \approx m_{\rm pl}$, implying $g<10^{-5}$. Clearly, this result implies that the required value of the coupling, $g$, to obtain efficient preheating is inconsistent with having a naturally light inflaton during inflation. In other words, in general it is expected that heavy $\chi$ fields running in the loops induced by the direct coupling $g^2\phi^2\chi^2$ tends to de-stabilize parameters of the inflationary sector if we insist on the effective particle production at the end of inflation.   

We have a good understanding of the limitations to the approximations we have used above to constrain preheating in chaotic inflation models, especially since these toy models have been well-studied over the years to establish when they lead to successful reheating.  At the same time, it is clear that we are seeing tension in analytic expectations for finding reliable preheating models.  It is also clear that doing a full non-linear analyses for all parameters in all models of preheating is not an efficient way to do model analysis.  Can one always establish a connection between the parameters during inflation and those same parameters during reheating?
 What is the expected mass of the reheat fields during inflation?
Can't the inflaton just decay through higher dimensional operators present at the time of reheating?
These are some of the questions we hope to address by developing a more systematic approach to reheating below.

\section{Reheating in Weinberg's Covariant formulation of the EFT of Inflation \label{sec2}}
In this section, we extend Weinberg's EFT approach to inflation \cite{Weinberg:2008hq} to include the end of inflation and the beginning of (p)reheating. Focusing on a two-field scalar field model for simplicity, we present both analytic and numeric results from our investigation into the background evolution and the resulting particle production.  We find that consistency of the background EFT within this approach limits its applicability and how well it can be used to successfully describe (p)reheating.  This will motivate us to consider a different approach in Section \ref{sec3}.

\subsection{Construction of the EFT}
Following \cite{Weinberg:2008hq} we consider the most general EFT of a scalar field in General Relativity which can be written as
\be \label{linf}
{\cal L}_{\rm inf}= -\frac{1}{2} m_{\rm pl}^2 R - \frac{1}{2} \left( \partial \phi \right)^2 - V(\phi) + \frac{c_1}{\Lambda^4}  \left(   \partial \phi \right)^4,
\ee
where $\Lambda$ is the UV cutoff of the theory, in general $c_1=c_1(\phi )$ is an arbitrary function of the scalar, and we have neglected terms involving the Weyl tensor which are suppressed relative to the leading correction \cite{Weinberg:2008hq}.  Assuming that the equations of motion admit inflationary solutions it was shown in \cite{Weinberg:2008hq} that this is also the most general EFT for the inflationary background (to be contrasted to the EFT for the perturbations which we will discuss in Section \ref{sec3}).

CMB observations imply that the power spectrum of scalar fluctuations is nearly scale-invariant, which can be realized through an approximate shift symmetry for the inflaton. 
This allows us to approximate $c_1(\phi)$ as nearly constant during inflation (its time evolution is slow-roll suppressed). 
When the EFT expansion is applicable, \ie $\Lambda > \dot{\phi}^{1/2}$, self-interactions of the inflaton are small and non-Gaussianity is negligible \cite{Maldacena:2002vr}. 

We now introduce an additional scalar that will play the role of the reheat field after inflation.
For simplicity, we will focus on the situation where the reheat field has an effective mass of at least the Hubble-scale during inflation to avoid considering multi-field inflation. However, the reheat field's mass during inflation 
is an important consideration which we comment on later.  Given these assumptions the starting point of our analysis is similar in spirit to that of \cite{Assassi:2013gxa}, where those authors considered the EFT of the inflationary background coupled to an additional scalar sector during inflation.
Again working to next-to-leading order in the derivative expansion we can introduce the Lagrangian for the additional scalar $\chi$,
\be \label{lchi}
{\cal L}_{\chi}= - \frac{1}{2} \left( \partial \chi \right)^2 - U(\chi) + \frac{c_2}{{\Lambda}^4}  \left(   \partial \chi \right)^4,
\ee
where $c_2$ and $U(\chi)$ are arbitrary functions of $\chi$, but can not contain the inflaton due to its approximate shift symmetry\footnote{The spontaneous or explicit breaking of the shift symmetry at the time of reheating can be important and creates an additional limitation of this approach.}.

Finally, we can introduce the interactions between the two sectors that respect the inflaton's shift symmetry -- implying that terms of the form $\phi^p \chi^q$ are forbidden.  At the level of dimension five operators it was shown in \cite{Assassi:2013gxa} that the shift symmetry can be used to forbid the operators $\partial_\mu \phi \partial^\mu \chi$ and $\chi \partial_\mu \phi \partial^\mu \chi$.  Similar arguments can be used at the level of dimension six operators and we find the two leading interactions\footnote{We have taken the cutoff of the EFT to be the same for both the inflationary and hidden sector for simplicity, although this need not be the case. We expect our main conclusions in this section to be insensitive to this assumption.}
\be \label{lmix}
{\cal L}_{\rm mix}= -c_3(\partial \phi)^2 \frac{\chi}{ \Lambda}  -  c_4(\partial \phi)^2 \frac{\chi^2}{\Lambda^2} + {\cal O}\left( \frac{1}{\Lambda^3}\right),
\ee
where $c_3$ and $c_4$ are expected to be order one constants and positive (for a UV completable EFT \cite{Adams:2006sv} and to avoid pathological instabilities \cite{Easson:2016klq}).
Given our discussion and assumptions above, the EFT of Inflation with an additional to-be reheat field is then given by, ${\cal L}= {\cal L}_{\rm inf} + {\cal L}_\chi + {\cal L}_{\rm mix}$.  
Focusing on the leading interactions we have
\be \label{goa}
{\cal L}=\frac{1}{2} m_{\rm pl}^2 R -\frac{1}{2} f\left(\frac{\chi}{\Lambda}\right) (\partial \phi)^2  -\frac{1}{2}  (\partial \chi)^2 -V(\phi) - U(\chi),
\ee
where
\beq\label{f}
 f\left(\frac{\chi}{\Lambda}\right) = 1 + 2c_3 \frac{\chi}{\Lambda} + 2c_4 \frac{\chi^2}{\Lambda^2}.
\eeq

The dynamics of fluctuations that arise from \eqref{goa} have been studied extensively in the context of inflation. In particular, there can be interesting signatures for both the power spectrum and higher point correlation functions (\eg non-gaussianity) depending on the mass of $\chi$ \cite{Chen:2009zp}, its stabilization \cite{Cremonini:2010ua,Achucarro:2010da,Tolley:2009fg,Shiu:2011qw,Avgoustidis:2012yc,Burgess:2012dz}, and whether the $\chi$ and $\phi$ sectors are strongly or weakly mixed \cite{Baumann:2011su}. 

In this work we are interested in connecting this system to the end of inflation and reheating. In particular, we would like to investigate if (p)reheating of the $\chi$ sector can be achieved through the derivative couplings in \eqref{f} as these are the leading interactions allowed by the shift symmetry of the inflaton. 

We note that (p)reheating with derivative couplings has been considered before. The authors of \cite{ArmendarizPicon:2007iv} have studied a particular realization of the EFT we are considering in this work.  In their case the approximate shift symmetry of the EFT resulted from a specific UV completion motivated by Natural Inflation 
\cite{Adams:1992bn}, where the spontaneous (and explicit) breaking of a $U(1)$ symmetry of a complex scalar resulted in an inflaton associated with the pseudo-Nambu-Goldstone Boson (pNGB) and the reheat field corresponded to the excitation of the radial direction. 
The UV theory took the form
\be \label{uvL}
{\cal L}=-(\partial_\mu \Phi)(\partial^\mu \Phi^*) - \lambda (F^2 - \Phi^*\Phi), 
\ee
where the $U(1)$ symmetry is broken by the vacuum solution $\langle | \Phi | \rangle = F$.  The inflaton potential results from the explicit breaking term
\be\label{VN}
V(\phi)=\mu^4 \left[ 1-\cos\left( \frac{\phi}{F} \right) \right].
\ee
Expanding around the vacuum solution using
\beq
\Phi = (F + \chi)~ e^{i\phi/F},
\eeq
one can easily see that this particular model can be recast as the EFT of the matter sector given by the Lagrangian \eqref{goa} with the replacement $\Lambda \to F$.  We note that in this particular class of models, adequate inflation unfortunately requires $F \gg m_{\rm pl}$, which seems to be at odds with additional non-perturbative corrections and expectations from quantum gravity \cite{ArkaniHamed:2003mz,Dimopoulos:2005ac}. 
However, we emphasize that the (bottom-up) EFT approach we are taking here is more general than this particular class of models. In particular, 
we emphasize (see also \cite{Assassi:2013gxa}) that the symmetries resulting in \eqref{goa}
may be the result of a fundamental symmetry of the UV theory (as in the example of \cite{ArmendarizPicon:2007iv}), but they can also be the result of an accidental symmetry in the IR, or the result of fine-tuning of the effective potential. In this way, the model of \cite{ArmendarizPicon:2007iv} provides a particular UV completion of the more general EFT approach we consider here. This is analogous to the way in which EFT methods can capture phenomenology near the scale of Electroweak symmetry breaking, without one having a precise description of the UV physics and mechanism responsible for breaking Electroweak symmetry.

In general, the inflaton potential $V(\phi)$ in our EFT is arbitrary and does not need to take the specific form given in \eqref{VN}.  We also have that the scale $\Lambda$ can be taken as $\Lambda < m_{\rm pl}$ without raising any immediate issues about the consistency of inflation. We will see the importance of this observation when we consider the dynamics of the background and fluctuations in the following sections.

\subsection{Analysis of Reheating in the EFT}

To justify using an EFT at the end of inflation, we need to ensure that the model is self-consistent, i.e. we have to check that there is a consistent background solution to the equations of motion for the fields,
\be
\label{phi}  \ddot{\phi}+3H \dot{\phi} + \partial_\chi\left( \ln f \right) \dot{\phi}\dot{\chi}+ f^{-1} \partial_\phi V=0,
\ee
and
\be
\label{chi}\ddot{\chi} + 3H \dot{\chi} - \frac{1}{2} (\partial_\chi f)~ \dot{\phi}_{{}}^2 +\partial_\chi U =0,
\ee
and that the background also admits a perturbative description.  This procedure will allow us to study the existence (or non-existence) of resonant phenomena, and establish when viable preheating occurs.

\begin{figure}[h!]
\begin{center} 
\includegraphics[scale=0.37]{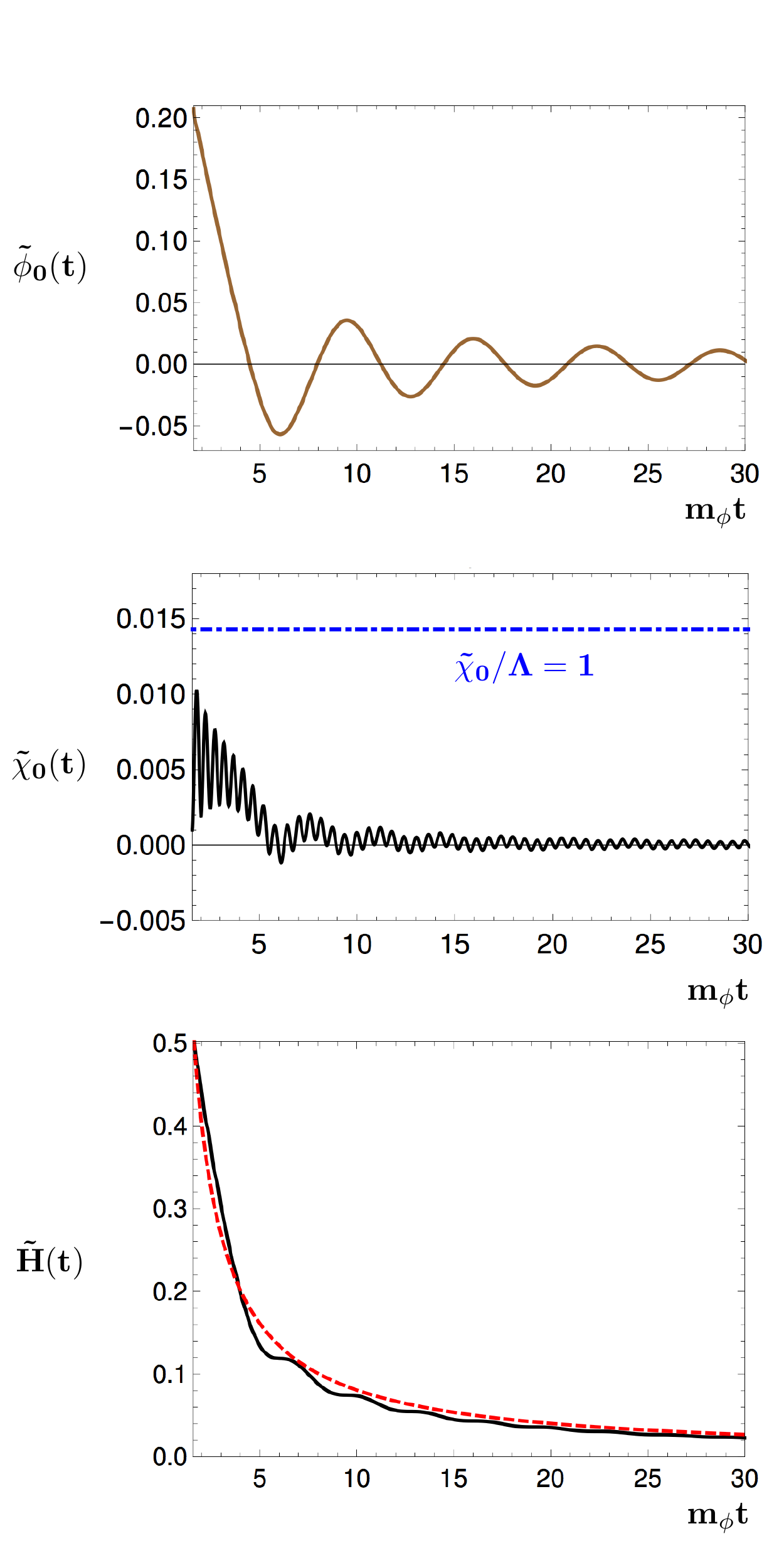}\\
\end{center}
\caption{ This figure gives the evolution of the background fields and Hubble parameter, where tildes imply we have normalized these quantities by $\sqrt{8\pi}m_{\mbox{\tiny pl}}$, and time is in units of the inflaton mass. 
For this realization we take $m_\chi/m_\phi = 10$, $m_{\rm pl} /\Lambda = 14$ and initial conditions $\phi_{_0} = 1.038~ m_{\rm pl}$, $\dot{\phi}_{_0}=-0.662~m_{\rm pl}$, $\chi_{_0}=\dot{\chi}_{_0} = 0.005 \; m_{\rm pl}$.  The top panel gives the evolution of the inflaton.
In the middle panel the solid black curve is $\tilde{\chi}_{_0}(t)$ and below the dot-dashed blue horizontal line marks the region where the  EFT of the background is valid. The bottom plot gives the Hubble rate where the red-dashed line represents a strictly matter dominated evolution. \label{fig:BG1}} 
\end{figure}  

We begin by studying the behavior of the background fields $\phi_{_0}$ and $\chi_{_0}$.  These are described by the following equations of motion,
\be \label{BG1}
\ddot{\phi}_{_0}+3H \dot{\phi}_{_0} + \partial_\chi\left( \ln f \right) \dot{\phi}_{_0} \dot{\chi}_{_0}+ f^{-1} \partial_\phi V=0,
\ee 
and
\be
\label{chigo} \ddot{\chi}_{_0}   + 3H \dot{\chi}_{_0} - \frac{1}{2} (\partial_\chi f)~ \dot{\phi}_{_0}^2 +\partial_\chi U =0.
\ee
If we further assume that the zero-mode dominates the energy density (and pressure) of the universe in the linear regime, then we can write down the evolution equations for the scale factor, 
\be \label{BG2}
H^2 = \frac{1}{3m_{\rm pl}^2}\left(\frac{1}{2}f  \dot{\phi}_{_0}^2  +\frac{1}{2} \dot{\chi}_{_0}^2 + V(\phi_{_0}) +U(\chi_{_0})\right),
\ee
and the Hubble parameter,
\be
\dot{H} = -\frac{1}{2m_{\rm pl}^2}\left( f \dot{\phi}_{_0}^2 + \dot{\chi}_{_0}^2\right).
\ee

The first question that we need to address is whether the zero-mode of the reheat field acquires a significant displacement from zero. Using \eqref{f}, and taking $c_3$ and $c_4$ to be order-one constants then \eqref{chigo} becomes  
\bea 
\label{chiz}
\ddot{\chi}_{_0} &+& 3H\dot{\chi}_{_0} + \partial_\chi U  - \frac{\dot{\phi}_{_0}^2}{\Lambda^2} \chi_{_0} - \frac{\dot{\phi}_{_0}^2}{\Lambda}=0,
\eea
The last two terms in \eqref{chiz} come from the EFT expansion -- \ie we have dropped terms in the Lagrangian of order $\sim \dot{\phi}_{_0}^2 \chi_{_0}^3 / \Lambda^3$ and higher.  Therefore, if either of these terms become large (e.g. if $\chi_{_0} / \Lambda > 1$) then the EFT expansion of the background is not justified.  Equation \eqref{chiz} is that of a harmonic oscillator with time-dependent frequency, where the last term resembles an external force, which we also require to be small compared to the restoring force from the effective potential.  Assuming that $U(\chi_{_0}) \approx m_\chi^2\chi_{_0}^2/2$, which is self consistent with our small-displacement assumption, we can find the stable minimum of the effective potential,
\beq
U_{\rm eff} = U(\chi_{_0}) -\fr{1}{2} \dot{\phi}_{_0}^2 f\left(\fr{\chi}{\Lambda}\right).
\eeq
to be  
\beq
\chi_{_0}(t) \simeq  \frac{\dot{\phi}_{_0}^2}{m_\chi^2 \Lambda} + {\cal O}\left(   \frac{\dot{\phi}_{_0}^4}{m_\chi^4 \Lambda^2}  \right).
\eeq   

The velocity of the inflaton at the end of inflation is roughly $\dot{\phi} \sim m_\phi m_{\rm pl}$, which allows us to write down an approximate condition on the size of $\chi_{_0}$, 
\be
\frac{\chi_{_0}}{\Lambda} < 1
\ee
implies that
\be
\label{key}\frac{m_\phi^2}{m_\chi^2} < \left(\frac{\Lambda}{m_{\rm pl}}\right)^2
\ee
That is, we find that we are free to lower the cutoff of the EFT below the Planck scale ($\Lambda \ll m_{\rm pl}$), but at the cost of increasing the mass of the reheat field above that of the inflaton. 
The fact that particle production is still possible in the $m_\chi \gg m_\phi$ regime emphasizes the importance of preheating versus reheating, since in this situation perturbative decays are kinematically forbidden.
It is also interesting that this condition is {\it independently} required so that the reheat field does not interfere with the the inflationary dynamics prior to reheating (constraints from non-Gaussianity could also be imposed).  That is, even for $m_\phi < m_\chi \simeq  3H_I$ such heavy fields can have a dramatic impact on inflation \cite{Chen:2009zp,Cremonini:2010ua,Achucarro:2010da,Tolley:2009fg,Shiu:2011qw,Avgoustidis:2012yc,Burgess:2012dz,Baumann:2011su}. 
We also note that the presence of a discrete $Z_2$ symmetry could be used to forbid the dimension five operator leading to the tadpole in \eqref{chiz}, and our stability condition \eqref{key} would still hold due to the presence of the dimension six operator.  

We have numerically verified the result \eqref{key} by solving the system \eqref{BG1}-\eqref{BG2} for a range of masses, initial conditions, and the cutoff $\Lambda$. 
In Figure \ref{fig:BG1}, we plot a particular realization of a consistent configuration for the background fields together with the evolution of the cosmological background. In the plot, we take $m_{\rm pl}/\Lambda = 14$ and $m_\chi/m_\phi =10$ consistent with \eqref{key}. We see that the background value $\chi_{_0}$ stays consistent within the EFT regime, while inflaton oscillations proceed as in the case of a quadratic potential. On the other hand, it can be seen that the expansion of the universe is slightly faster than $H(t) \propto t^{-1}$ initially, and then asymptotes to this behavior at late times $m_\phi t \gg 1$. 
We conclude this section by emphasizing that in order to have a stable, well-behaved background solution within the regime of validity of the EFT, one requires the condition, \eqref{key} to be satisfied. 

\subsubsection{Non-perturbative Dynamics and Limitations of the Background EFT}
We now consider whether resonant particle production is possible around the background we analyzed in the previous section.
Expanding both scalar fields to first order around their background values, $\phi = \phi_{_0} + \delta\phi$, $\chi = \chi_{_0} + \delta\chi$ in the Lagrangian \eqref{goa}, we write the equation of motion for the linearized fluctuations of the reheat field in Fourier space as
\beq\label{mchi}
\delta\ddot{\chi}_{k} + 3H\delta\dot{\chi}_k + \left[ \left( \frac{k}{a}\right)^2 +m_{\chi}^2 - \frac{ \dot{\phi}_{_0}^2}{\Lambda^2} \right] \delta\chi_k=  2 \frac{\dot{\phi}_{_0}}{\Lambda}\left[1+\frac{\chi_{_0}}{\Lambda}\right]\delta\dot{\phi}_k,
\eeq
where the terms on the right side are due to the mixing with inflaton fluctuations. These terms can source $\delta\chi_k$ fluctuations whenever $\delta\dot{\phi}_k$ is large. In the initial stage of (p)reheating the effect of this term will be negligible.  Neglecting these terms, we focus on sub-Hubble scales first neglecting the cosmological expansion (we take $a(t) \to 1$, $H(t) \to 0$).   In this approximation, \eqref{mchi}  becomes
\beq
\delta\ddot{\chi}_{k} + \left[ k^2 + m_{\chi}^2 - \frac{ \dot{\phi}_{_0}^2}{\Lambda^2} \right] \delta\chi_k = 0,
\eeq
where we define the frequency of the modes as $\omega_k ^2 (t) = k^2 +m_\chi^2 -\dot{\phi}_{_0}^2/\Lambda^2$.
Given a coherently oscillating inflaton, $\phi_{_0} = \Phi (t)\sin(m_\phi t) $, we can map this mode equation to the Mathieu equation
\beq\label{Mathieu}
\delta{\chi}_k'' + \left[A_k - 2q \cos(2z)\right]\delta\chi_k = 0,
\eeq
where we have defined the dimensionless time $z = m_\phi t$ and $A_k = (k^2 + m_\chi^2)/m_\phi^2 - 2q$ with $q = \Phi^2/4\Lambda^2 $. Floquet's theorem \cite{VogtNilsen:1956mr} states that for a given wave-number, \eqref{mchi} has solutions of the form 
\beq
\delta\chi_k = e^{\mu_k z} g_1(z) + e^{-\mu_k z} g_2(z) ,
\eeq 
where $g_1$ and $g_2$ are periodic functions and $\mu_k$ is the Floquet exponent. In general, the Floquet exponent $\mu_k$ depends on the wave number $k$, the mass of the reheat field $m_\chi$, and the ratio $\Phi/\Lambda$. For cases where the real part of the exponent is non-zero, we have exponentially growing modes of $\delta\chi_k$. 

\begin{figure}[t!]
\begin{center} 
\includegraphics[width=1.0\columnwidth]{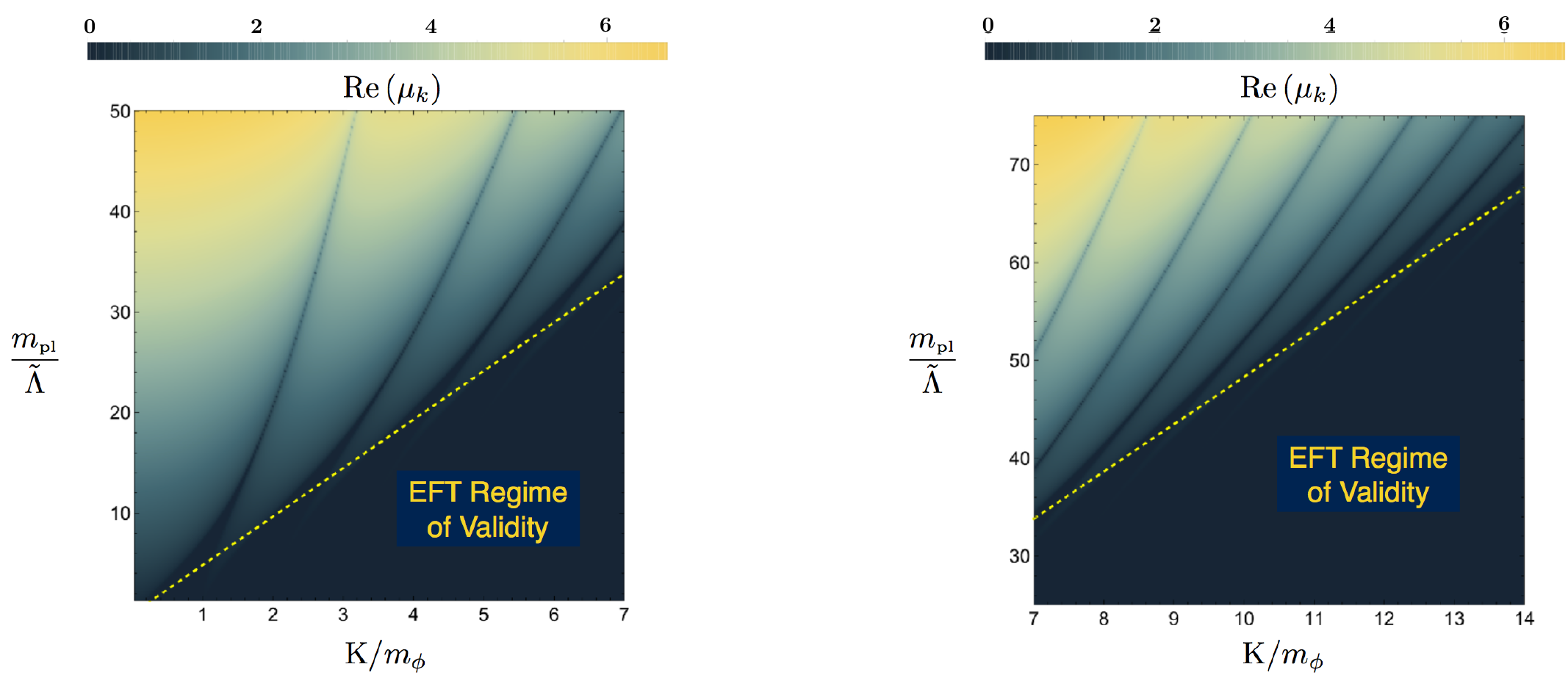}
\end{center}
\caption{Instability band structure for the model $V_{\rm tot}= \fr{1}{2}m_\phi^2 \phi^2 + \fr{1}{2}m_\chi^2 \chi^2 -\fr{1}{2} \dot{\phi}_{_0}^2 f\left(\fr{\chi}{\Lambda}\right)$, where $f$ is given by \eqref{f}. This density plot represents the real part of the scaled Floquet exponent, $\mbox{Re}(\mu_k)$, where lighter regions represent larger values. The y-axis is the hierarchy between the Planck mass and the rescaled cut-off of the EFT, $\tilde{\Lambda}=\Lambda / \sqrt{8\pi}$, while the x-axis corresponds to $K=\sqrt{k^2 +m_\chi^2}$ in units of $m_\phi$.   \label{fig:Mathieu} } 
\end{figure} 

The structure of \eqref{Mathieu} tells us that the resonant momenta are grouped into bands in parameter space.  Since $k^2 > 0$, and hence, $A_k > -2q$, there are also meaningful statements one can make about the regions of the Mathieu parameter space that are probed by our reheating models.   One interesting case is when some modes satisfy $-2q < A_k < 0$;  in this case, \eqref{Mathieu} assures us that there's a time when the mass-squared of the these modes is negative (analogous to the cases explored in \cite{Greene:1997ge}) and the Floquet exponent can be very large, $\mu_k \simeq (4/\pi)~ q^{1/2}$ for $q \gg 1$. There's another case in which $0 < A_k < 2q$, where the mass-squared of some of the $\delta\chi_k$ modes become tachyonic for certain time intervals and is also very efficient (analogous to \cite{Felder:2000hj}.) 

On the other hand, $A_k$ is frequently larger than $2q$.  While these models have parametric instabilities, the resonance structure requires us to be more careful.  For our purposes here, the consistency of the background EFT requires the mass of the reheat field to satisfy $m_\chi^2 > \dot{\phi}_{_0}^2/\Lambda^2$, which requires avoiding the regions of the parameter space that guarantee strong, broad, resonance. While the inflaton undergoes periodic oscillations this condition implies 
\beq\label{key1}
m_\chi^2 > m_\phi^2 \fr{\Phi^2}{\Lambda^2},
\eeq
which is exactly what we have found in equation \eqref{key} with $\Phi = m_{\rm pl}$. Here, we have used $\phi(t) =\Phi \sin (m_\phi t)$ considering the maximum value of $\dot{\phi}_{_0}^2/\Lambda^2$.  
We have also studied this system numerically, using FloqEx \cite{floqex}, with our results appearing in Figure \ref{fig:Mathieu}.  The figure shows the magnitude of the Floquet exponent as a function of cutoff and wave number.  One can see the broad (and tachyonic) resonance regimes mostly live outside of those probed by the EFT. We must keep in mind, though, that these estimates could still produce some particles through parametric resonance, and should be studied through full lattice methods -- we leave this to future work.

Our main conclusion thus far is that if we require the reheat field to respect the shift symmetry of the inflationary sector (implying adequate inflation consistent with CMB observations), successful reheating suggests considering an EFT cutoff far below the Planck scale $\Lambda \ll m_{\rm pl}$.  We saw that having such a sub-Planckian cutoff can quickly lead to the breakdown of the background EFT expansion when we require 
efficient reheating in the EFT. 

As another example of when the EFT expansion may breakdown, consider the corrections we have thus far neglected in \eqref{linf}. When evaluated on the background the operator contains a term
\be
\frac{c_1}{\Lambda^4}  \left(   \partial \phi \right)^4 \supset  \frac{c_1}{\Lambda^4}  \dot{\phi}_0^2 \left(   \partial \phi \right)^2.
\ee
During inflation this term will be slow-roll suppressed $\dot{\phi}_0^2 / \Lambda^4  \sim \epsilon \, H^2 m_{\rm pl}^2 / \Lambda^4$ and higher order terms will be even further suppressed as long as $\Lambda$ is not far below $m_{\rm pl}$ during inflation\footnote{Using the power spectrum normalization one can also show the condition $\dot{\phi}_0^2 / \Lambda^4 <1$ implies a lower bound $\Lambda / m_p \gtrsim \sqrt{\epsilon} \, 10^{-2}$, where $\epsilon=d(H^{-1})/dt$ is the slow-roll parameter.}. However, for smaller values of the cutoff this corresponds to strong coupling of the background and our EFT approach breaks down -- this would also lead to a large level of non-Gaussianity \cite{Baumann:2011su}. Assuming the background remains weakly coupled at the end of inflation we have
\be \label{constraint}
\frac{c_1}{\Lambda^4}  \dot{\phi}_0^2  \sim \frac{m_\phi^2 \phi_e^2}{\Lambda^4}
\sim \left(  \frac{m_\phi}{m_{\rm pl}} \right)^2 \left( \frac{\phi_e}{m_{\rm pl}} \right)^2 \left( \frac{m_{\rm pl}}{\Lambda} \right)^4,
\ee
so for $\Lambda$ far below the Planck scale the EFT would again fail as this term would be as important as the kinetic term (and terms even higher in derivatives that we neglected would also be important). For example, in chaotic inflation where the inflaton mass is fixed by the COBE normalization this implies $\Lambda \gtrsim 10^{-3} \, m_{\rm pl}$.  
We emphasize that this constraint has nothing to do with requiring adequate inflation and is an added constraint for the consistency of the derivative expansion of the EFT during reheating.  We now turn to a different EFT approach where the challenges discussed in this section can be addressed.

\section{The EFT of (p)Reheating \label{sec3}}
\begin{figure}[h]
\begin{center} 
\includegraphics[width=0.60\columnwidth]{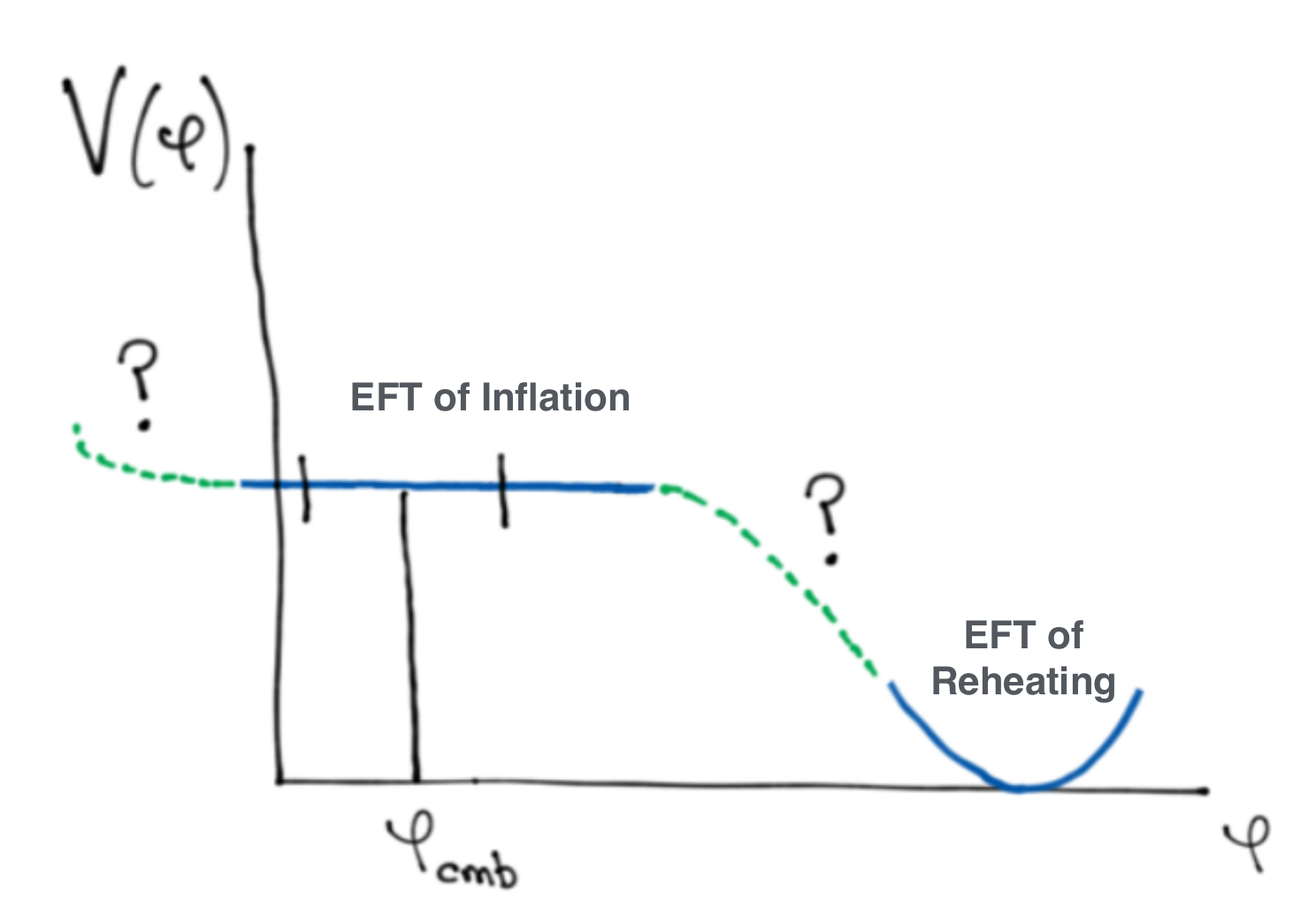}
\end{center}
\caption{Obtaining adequate inflation, ending inflation and then successful reheating in the EFT requires a complete knowledge of the inflationary potential.  This presents a challenge when using Weinberg's EFT approach to capture reheating in many classes of models. \label{figcartoon}} 
\end{figure}
We have seen that using an EFT approach to the background has limited utility in simultaneously describing inflation and reheating.
Indeed, in addition to the challenges discussed at the end of Section \ref{sec2}, an additional concern is that there could be terms that badly break the shift symmetry at the time of reheating. Such terms could be small during inflation (suppressed by the breaking scale), but could be important at the time of reheating.  Alternatively, there are many reheating models in which the shape of the potential during inflation is vastly different than it is during reheating (and could include additional fields like in hybrid models) and the background EFT approach requires a knowledge of the complete potential. This is illustrated in Figure \ref{figcartoon}. 
 
In particular, the terms arising from the breaking of the shift symmetry of the inflaton (which would include thus far forbidden terms of the form  $g_i \phi^p \chi^q$) could become as important as the other terms we have considered 
in \eqref{mchi}.  As another example consider
the potential 
\be \label{mustafa}
V=\frac{m^2 M^2}{2 \alpha} \left[ \left( 1+\frac{\phi^2}{M^2} \right)^\alpha - 1\right].
\ee
where $\alpha<1$.  This toy model captures many important inflationary models including axion monodromy \cite{Amin:2011hj}.  During the inflationary phase this potential scales as $V \sim \phi^{2\alpha}$ and is sensitive to the scale $M$, whereas the behavior during reheating ($\phi < M$) is independent of $M$ and
$V \sim m^2 \phi^2$.  So in our EFT approach expanding the field in powers of $\phi/\Lambda$ is causing us to miss these types of theories.

In addition, new degrees of freedom could appear at the time of reheating that were heavy during inflation and could have been integrated out -- in other words the EFT during inflation and the EFT during reheating can correspond to two distinct EFTs.  This is not to say our approach doesn't capture many models.  
In particular, we've seen that the model of \cite{ArmendarizPicon:2007iv} is captured by our approach, and most chaotic inflation models would be as well. But even focusing only on the inflationary epoch we know that Weinberg's EFT is not capable of capturing a large number of interesting models.  For example, in DBI type models where the background is in some sense strongly coupled one needs a non-perturbative expression for the background as it is a resummed expression where each derivative in the derivative expansion must be kept, e.g. $V \sim \sqrt{1-\dot{\phi}^2/\Lambda^4}$.  Such models are not captured by the Lagrangian of \eqref{linf}.  One may also anticipate reheating models where the background of the reheat field could also exhibit such non-linear behavior and then the derivative expansion of the Lagrangian \eqref{lchi} would be inadequate -- as well as the expansion of the mixing terms stopping at dimension six in \eqref{lmix}.
One final objection is that we have only concentrated on scalar reheat fields.  Reheating to fermions and gauge fields is also important, and the way in which this proceeds is not only model dependent, but the spin statistics can also make important differences in the efficiency of reheating \cite{Adshead:2016xxj}.

Given these shortcomings of the EFT of the background we now turn to construct an EFT for reheating along the lines of the EFT of Inflation \cite{Cheung:2007st}.  As we will discuss, this approach can overcome many of the obstacles established in this Section.  In the remainder of this section, we first begin by constructing an EFT focusing on the fluctuations directly at the end of inflation.  This theory will share many similarities with the EFT of multi-field inflation \cite{Baumann:2011su,Senatore:2010wk}. However there will be important differences which we will discuss.  We then demonstrate how the approach can reproduce both the results of self resonant reheating and multi-field reheating.  We also discuss some new models that arise from considering the symmetries of the EFT.

\subsection{Construction of the EFT of Fluctuations}
The EFT expansion in fluctuations (rather than the background) relies on the fact that the background expansion of the universe spontaneously breaks time-translation invariance. 
Over the history of the universe there have been many different dominant forms of matter and energy, and so many different sources of time-translation breaking including; inflatons, post-inflation / pre-BBN fields, radiation, dark matter, and eventually dark energy today.  As the universe passed through these phases the energy density changed its composition many times, but the scale factor continued to monotonically increase.  The EFT approach takes this background evolution as given {\it a priori} (as specified by the background functions $a(t)$, $H(t)$, and $\dot{H}(t)$) and focuses directly on the most general EFT for the fluctuations around this background. 

In taking this approach we give up on realizing explicit models for the background, and instead focus on implications and observations associated with the fluctuations.  In regards to connecting with observations this approach is adequate\footnote{Although the connection to observables is not necessarily always straightforward \cite{Linder:2015rcz}.}, since physical observables correspond to fluctuations and not background quantities \cite{Cheung:2007sv}. The approach also has the advantage 
that the underlying physics responsible for driving the background expansion can be non-perturbative, in the sense that the background doesn't need to admit an EFT expansion (as we required in Section \ref{sec2}).  
Instead, this EFT approach is more general and models are classified by their symmetry breaking properties and the allowed operators in the Lagrangian correspond to cosmological perturbations.  In many cases the symmetries alone can be used to establish rigid constraints on the theory of the fluctuations and associated observables.  For example, it is well known that inflation requires that 
de Sitter symmetry must be non-linearly realized and this leads to constraints on inflaton correlation functions.  This fact is manifest in the EFT of Inflation approach using the corresponding Goldstone boson \cite{Cheung:2007sv}.  
This EFT approach has also been shown to be useful when the cosmological background changes its behavior, e.g. in the EFT of dark energy \cite{Gubitosi:2012hu,Bloomfield:2012ff,Fasiello:2016qpn,Fasiello:2016yvr,Lewandowski:2016yce}, where one is primarily interested in observations during matter domination, but also must account for observations during dark energy domination.

The generality of the EFT approach when applied to cosmological backgrounds was first established in  \cite{Creminelli:2006xe}, where the 
authors were investigating violations of the Null Energy Condition in non-standard cosmologies.
In that paper, referencing earlier work of Weinberg \cite{Weinberg:2003sw}, it was pointed out that on long wavelengths there is always an adiabatic mode corresponding to the Goldstone boson of spontaneously broken time diffeomorphism invariance.  Whenever a decoupling limit exists -- in which the Goldstone decouples from gravity -- this broken symmetry is then realized as spontaneously broken time translation invariance (the gauge symmetry effectively becomes a global symmetry). Thus, for any FRW spacetime it is possible to utilize the EFT approach and it is in this vain that we will construct our EFT for reheating following the initial ideas presented in \cite{Ozsoy:2015rna}. 

As an example, suitable for studying the dynamics at the end of inflation, we can consider a decelerated FRW expansion with the background metric
\beq
ds^2 = - dt^2 + a^2(t) \delta_{ij} dx^i  dx^j, ~~~ \ddot{a}(t) < 0.
\eeq
We can think of this background as generated by the evolution of a set of homogeneous scalars\footnote{In general, we are not restricted to scalar fields, \eg another example can be a set of perfect fluids.} fields, \ie $\{\phi_{_0},\chi_{_0},\dots\}$. In this work, to study dynamics at the end of inflation, we may consider only one of the scalars, \eg the inflaton $\phi_{_0}$, that contributes significantly to the evolution of scale factor, $a(t)$. This FRW evolution has a preferred time slicing described by the homogeneous scalar which can also be considered a clock. In order to describe the theory of fluctuations around this background, we can go to a co-moving frame (unitary gauge)
where the vacuum expectation value of the scalar coincides with this privileged time slicing, corresponding to distinct values of $\langle\phi \rangle = \phi_{_0}$. As we have fixed the slicing of space-time, general time diffs\footnote{As we mentioned before our main interest is the global part of time diffs, \ie time translations. See \cite{Piazza:2013coa} for more discussion on this matter.} are no longer a symmetry and the fluctuations of the scalar are hidden in the metric perturbations, which now describe three degrees of freedom: two transverse for the graviton and one for the scalar. We can always re-introduce inflaton fluctuations by a common local shift in time, \ie $t \to t + \pi(x)$. By definition, such a fluctuation corresponds to an adiabatic fluctuation, proportional to Goldstone mode $\delta \phi = \dot{\phi}_{_0}\pi$ associated with the broken symmetry. In this work, apart from the adiabatic fluctuations, we will consider an additional degree of freedom $X(t,x)=\chi_{_0}(t)+\chi(t,x)$, which will play the role of the (p)reheat field. As is standard in the literature we will take this field 
to be a subdominant source of background evolution during the first stages of preheating (\ie $\rho_\phi \gg \rho_\chi$) since before particle creation $\langle X \rangle \simeq 0$. 
\subsubsection{The Action in Unitary Gauge}
The procedure for constructing the EFT of fluctuations for the inflationary sector coupled to a reheat field {\it at the time of reheating} is similar to the case of quasi-single field inflation considered in \cite{Noumi:2012vr}. Those authors considered the effects of particle production during inflation, whereas here we consider reheating and important differences will be discussed below.  
Nevertheless, the action can be constructed analogously and working in unitary gauge the action for the fluctuations is
\be
\label{theaction}
S=\int d^4 x \sqrt{-g}\left[\frac{m_{\rm pl}^2}{2}R-f_1(t)-f_2(t)g^{00}+ F^{(2)}(\delta g^{00},\chi,\delta R_{\mu\nu\rho\sgm},\delta K_{\mu\nu};\nabla_{\mu};t)\right],
\ee 
where $f_1$ and $f_2$ are arbitrary functions of time, $F^{(2)}$ starts quadratic in operators which must be covariant in spatial indices but not in time, $\nabla_{\mu}$ is the covariant derivative, and $\delta R_{\mu\nu\rho\sgm}$ and $\delta K_{\mu\nu}$ are the fluctuations in the Riemann tensor and extrinsic curvature, respectively. Note that the second and third terms in the above action are the only ones that contain linear perturbations. Requiring that terms linear in the fluctuations vanish (\ie tadpole cancelation)
follows from enforcing the background equations of motion in an FRW background \cite{Cheung:2007st},
\beq
3H^2 m_{\rm pl}^2 = f_1(t)+f_2(t),
\eeq
and
\beq
-2\dot{H}m_{\rm pl}^2= 2f_2(t).
\label{background_eom}
\eeq
As a simple example of tadpole cancelation, consider the end of inflation where the inflaton begins oscillating with a potential $V(\phi)$ and where derivative interactions and the density of other fields are negligible. In this case the functions in \eqref{background_eom} are given by $f_1= V(\phi_{_0})$ and $f_2=\dot{\phi}_{_0}^2/2$.  However, more generally, $f_1$ and $f_2$ can take any form as long as the background corresponds to the (p)reheating period, \ie an FRW universe with possibly small corrections due to oscillations.  For example, we could have a preheating model corresponding to DBI-like models of inflation where a large number of derivative self-interactions could play an important role \cite{Child:2013ria}. In that case the functions $f_1$ and $f_2$ would contain terms with an infinite number of derivatives at the level of the background. The key is that the behavior of the matter sector will be captured by the functions $f_1$ and $f_2$, and once we cancel the tadpoles, the background is then given (by the equations of motion) by $H(t)$ and its derivatives. Then, we can focus on the EFT of the fluctuations about this background -- just as in the case of the EFT of Inflation or DE \cite{Cheung:2007st,Bloomfield:2012ff,Gubitosi:2012hu}.
Thus, the problem we encountered in the previous section, where we would need to keep all the terms in the $\chi/\Lambda$ expansion is not an issue here.  
Instead, these terms are captured by $H$ and $\dot{H}$ and could represent re-summed, non-perturbative expressions for the background\footnote{The importance of strong coupling and resummation appears in many areas of physics including QCD and theories of modified gravity. See e.g. \cite{deRham:2014zqa}. An approach to strong coupling during preheating, using methods of holography, appeared in \cite{Cai:2016lqa,Cai:2016sdu}.}.  Moreover, because we are {\it not} performing a perturbative expansion of the background, we work under the assumption that we have a complete knowledge of the potential overcoming the problems associated with Figure \ref{figcartoon}.

The most general action is found by expanding the function $F^{(2)}$ in \eqref{theaction} in terms of fluctuations $\{\delta g^{00}, \chi,\delta K_{\mu\nu},\delta R_{\mu\nu\rho\sgm}\}$ and their derivatives. We emphasize that this EFT expansion is one in perturbations and derivatives. During reheating, the fluctuations are also assumed to be initially small, however significant particle production can change this (as we will discuss). Whereas the derivative expansion follows from locality, causality and unitarity in an FRW universe. 
In the gravity sector, $\delta g^{00}$ is a zero derivative object, whereas $\delta K_{\mu\nu}$ corresponds to one derivative and $\delta R_{\mu\nu\rho\sgm}$ to two, as they contain first and second order derivatives of the metric, respectively. When we introduce the Goldstone boson in the next section, it will be clear that terms with $\delta K$ and $\delta R$ will include higher derivatives of the Goldstone boson. Finally, we find it convenient to split the action in \eqref{theaction} into three parts
\beq
S = S_{\rm g} + S_\chi + S_{\rm g \chi},
\eeq
where the action $S_{\rm g}$ contains only terms build out of $\{\delta g^{00},\delta K_{\mu\nu},\delta R_{\mu\nu\rho\sgm}\}$, $S_\chi$ contains those purely from $\chi$ and the action $S_{\rm g \chi}$ is due to mixing between gravity sector and $\chi$. Following our discussion above, we then have 
\be
\label{SU} S_{\rm g}=\int d^4x \sqrt{-g} \left[\frac{m_{\rm pl}^2}{2}R - m_{\rm pl}^2\left(3H^2(t)+\dot{H}(t) \right) + m_{\rm pl}^2\dot{H}(t)g^{00} 
+ \frac{m_2^4(t)}{2!} \left( \delta g^{00} \right)^2 +\ldots \right],
\ee
\be
\label{Ss} S_{\chi}=\int d^4x\sqrt{-g}\left[-\frac{\alpha_1(t)}{2}g^{\mu\nu}\partial_\mu\chi\partial_\nu\chi 
+\frac{\alpha_2(t)}{2}(\partial^0\chi)^2-\frac{\alpha_3(t)}{2}\chi^2+\alpha_4(t)\chi\partial^0\chi\right], \;\; 
\ee
\be
\label{Sm}  S_{{\rm g}\chi} =\int d^4x\sqrt{-g}\left[\beta_1(t)\delta g^{00}\chi+\beta_2(t)\delta g^{00}\partial^0\chi+\beta_3(t)\partial^0\chi-(\dot{{\beta}}_3(t)+3H(t)\beta_3(t))\chi\right], 
\ee
where $g^{00}=-1+\delta g^{00}$ and the dots represent terms higher order in fluctuations and derivatives. Here, $\{m_2(t),\alpha_i(t),\beta_i(t)\}$ are thus far arbitrary functions of time that are permitted in the unitary gauge as time diffs have been spontaneously broken by the background. 
We note that the coefficient of the $\delta g^{00}$ operator is fixed by the background, implying that it is {\it universal} in the sense that all preheating models with the same background evolution will have the same coefficient (specified by $H(t)$ and its derivatives). Whereas, the operator $\left( \delta g^{00} \right)^2$ is an example of a {\it non-universal} 
operator, because $m_2$ is not fixed by the symmetries of the FRW background. Instead its value corresponds to a specific class of models (those with a non-unity sound speed). Similarly, broken time diffs generally allow for a term proportional to $\alpha_2$ that leads to non-trivial sound speed $c_\chi = \alpha_1/ (\alpha_1 +\alpha_2)$ in the reheat sector $\chi$. In \eqref{Sm}, the functions $\beta_i$ can be seen as a measure of the strength of mixing with gravitational fluctuations (including one scalar d.o.f). At this stage, the usefulness of this approach might be in question, given the large number of free parameters.  However, as we will see in the following sections, even though this is the most general theory to quadratic order, in practice many of the terms in \eqref{SU}-\eqref{Sm} are not important for elementary processes within reheating. 
Finally, we can further simplify the action by performing a field re-definition of $\chi$, using that $\chi=0$ on the background trajectory and using time reparametrization invariance to set $\alpha_4 =\beta_3=0$
in the actions \eqref{Ss} and \eqref{Sm}. 

The form of \eqref{SU}, \eqref{Ss} and \eqref{Sm} are not particularly useful in studying the dynamics as the scalar fluctuation representing inflaton is not manifest. We can re-introduce diffeomorphism invariance and the Goldstone mode related to inflaton by the St\"uckelberg trick, which will be our main focus in the following section.

\subsubsection{Introducing the Goldstone Boson}
To introduce the Goldstone boson along with time diffs, we first perform the broken time diffs $t\rightarrow t+\xi^0(t,\vec{x})$ in the actions \eqref{SU}-\eqref{Sm}. Since the cosmological background (\ie $H,\dot{H}$) as well as the free functions $\{\alpha_i,\beta_i\}$ depend on cosmic time, $t$.  The gauge function, $\xi^0$, will appear explicitly in the actions for the perturbations. 
We then replace $\xi^0\rightarrow \pi(t,\vec{x})$ everywhere it appears in the action and require that the Goldstone transforms non-linearly, $\pi \rightarrow \pi - \xi^0$ under diffs. In this way, clearly full diffeomorphism invariance can be restored in \eqref{SU}-\eqref{Sm}. In order to find the explicit form of the actions including the Goldstone $\pi$, we need to know the transformation rule for the remaining operators appearing in \eqref{SU}-\eqref{Sm} under $t\to t + \pi$.
Under the transformation we have
\bea\label{transform}
g^{00}&\rightarrow&g^{00}+2g^{0\mu}\partial_\mu \pi + g^{\mu \nu} \partial_\mu \pi \partial_\nu \pi \nn, \\
g^{i0}&\rightarrow&g^{i0}+g^{i \nu} \partial_\nu \pi \nn, \\
\partial^{0}\chi &\rightarrow& \partial^{0}\chi +g^{\mu \nu} \partial_\mu \chi\partial_\nu \pi \nn, \\
f(t) &\rightarrow& f(t+\pi) \\
 R_{\mu \nu \lambda \sigma} &\rightarrow& R_{\mu \nu \lambda \sigma}\\
\int d^4 x \sqrt{-g} &\rightarrow& \int d^4 x \sqrt{-g}
\nn
\eea
where $f(t)$ represents any time-dependent function appearing in the action. Carrying out this procedure on the action \eqref{SU} we find
\begin{align} \label{Sgpi}
S_{\rm g} &=\int d^4x \sqrt{-g}\Bigg[\frac{m_{\rm pl}^2}{2}R-m_{\rm pl}^2\left(3H(t+\pi)^2+\dot{H}(t+\pi)\right) \nn\\
\nn&~~~~~~~~~~~~~~~~~~~~+m_{\rm pl}^2\dot{H}(t+\pi)(g^{00}+2g^{0\mu}\partial_\mu\pi+g^{\mu\nu}
\partial_\mu\pi\partial_{\nu}\pi)\\ 
&~~~~~~~~~~~~~~~~~~~~+\fr{m_2^{4}(t+\pi)}{2!}(\delta g^{00}+2g^{0\mu}\partial_\mu\pi+g^{\mu\nu}
\partial_\mu\pi\partial_\nu\pi)^2 \Bigg]\ .
\end{align}
We see that this action is invariant under time diffs if we require the Goldstone to transform as $\pi \rightarrow \pi - \xi^0({t,\vec{x}})$, \ie the symmetry is non-linearly realized \cite{Senatore:2010wk}.  
We also note that requiring the symmetry be realized in the UV has forced relationships between the various operators (all the terms in parentheses must have the same coefficients). Following the same steps, \eqref{Ss} and 
\eqref{Sm} become 
\begin{align}
\nn S_{\chi}&=\int d^4x\sqrt{-g}\left[-\frac{\alpha_1(t+\pi)}{2}g^{\mu\nu}\partial_\mu\chi\partial_\nu\chi+\frac{\alpha_2(t+\pi)}{2}(\partial^0\chi+\partial_\mu\pi \partial^\mu\chi)^2  \right.  \\  \label{Sspi} &\left.~~~~~~~~~~~~~~~~~~~-\fr{\alpha_3(t+\pi)}{2}\chi^2\right],\\
\nn  S_{{\rm g}\chi} &=\int d^4x\sqrt{-g}\Big[\beta_1(t+\pi)(\delta g^{00}+2\partial^0\pi+\partial_\mu\pi\partial^{\mu}\pi)\chi \\
&~~~~~~~~~~~~~~~+\beta_2(t+\pi)(\delta g^{00}+2\partial^0\pi+\partial_\mu\pi\partial^{\mu}\pi) 
 (\partial^0\chi+\partial_\mu\pi \partial^\mu\chi)\Big]\label{Smspi}.
\end{align}
Similar to the discussion above, the non-linearly realized symmetry introduces interactions between $\chi$ and the Goldstone, $\pi$.

To describe the dynamics at the end of inflation, working with the full action given by $S_{\rm g}+S_{\chi}+S_{\rm g \chi}$ in complete generality is a difficult task. First of all, we need to have some input for the time-dependent functions, \ie $\{H(t),\alpha_i(t), \beta_i(t)\}$ appearing in the Lagrangian. However, as we will see, an investigation on the background dynamics during reheating along with the associated symmetries and scales of interest will allow us to obtain generic information on the form of these functions. This will be our main focus in the next section.

\subsection{Background Evolution During Reheating and Symmetries of the Action}\label{BevS}
\subsubsection{Background Evolution and Symmetries}
In parametrizing the background expansion we have assumed a decelerating FRW universe. A simple example is provided by a perfect fluid with an equation of state $w$ and with corresponding scale factor $a(t)\propto t^{2/3(1+w)}$ and expansion rate $H(t)=\dot{a}/a \propto t^{-1}$ with $H^{-1}$ setting the cosmic time scale.
On the other hand, in studies of the dynamics at the end of inflation the frequency of inflaton oscillations introduce another important time scale. 
For example, if the inflaton oscillates in a power-law potential, $V\propto \phi_{_0}^n$, the period of oscillations will be $2\pi\omega^{-1}=4 \int_0^{\phi_i}  \d\phi_{_0} ~(V(\phi_i) - V(\phi_{_0}))^{-1/2}$, which for general $n$ depends on the initial amplitude, $\phi_i$ \cite{Johnson:2008se}. 
In the limit that the period of oscillations is much smaller than the expansion time scale, $\omega^{-1} \ll H^{-1}$, coherent scalar field oscillations behave like a perfect fluid with an average equation of state, $\langle w \rangle_{\rm a} = (n-2 )/ n+2$ \cite{Turner:1983he}.  

The presence of two different time scales leads to  interesting symmetry breaking patterns within the EFT, and whether a symmetry is realized will depend on the dynamics under investigation.
At high energies (or small wavelengths) the energy being probed $E_{\rm{probed}}$ exceeds both the oscillation and expansion energy \ie $E_{\rm{probed}} \gg \omega \gg H$ and so the time evolution of the oscillator and the cosmic expansion is negligible -- time-translations are a good symmetry.
As we lower the energy scale to $E_{\rm{probed}} \lesssim \omega$ we first break time-translation invariance down to a discrete symmetry $t \to t + 2\pi \omega^{-1}$.  Then as we further lower the energy to 
$E_{\rm{probed}} \lesssim H \ll \omega$ this discrete symmetry is further broken by the cosmic expansion.  This symmetry breaking reflects that   
on large scales (low-energy) we have an expanding universe, but on sub-Hubble scales the only time dependence results from the oscillating scalar field and the effect of the expansion can be ignored. And at even higher energies (smaller distances / faster time scales) the scalar oscillations would not be probed. 

This hierarchy in scales can be captured by parameterizing the background behavior by a Hubble rate that is a sum of a monotonically evolving part and a small rapidly oscillating component,
\be \label{H_osc}
H(t) = H_{\frw}(t) + H_{\rm osc}(t) P(\omega t),
\ee
where the first term is adiabatically evolving $H_{\frw}(t)\propto t^{-1}$ and monotonically decreasing, whereas the second term leads to an oscillatory correction described by a general periodic function $P(\omega t)$ with period $T = 2\pi\omega^{-1}$. In order to ensure an overall monotonic FRW evolution we take the first term to be dominant, $H_{\frw} \gg H_{\rm osc}$. This implies our clock is always monotonically increasing -- as exemplified by the monotonic evolution of the scale factor $a(t)$ in an FRW universe.  This situation is to be contrasted with models where the universe itself is oscillating \cite{Bains:2015gpv}, which can exhibit a number of pathologies \cite{Easson:2016klq}. We also note that the time dependence of $H_{\frw}$ and $H_{\rm osc}$ is slow compared with the time scale of oscillations $\omega^{-1}$, \ie $\dot{H}_{\frw}/(H_{\frw} \omega)\sim \dot{H}_{\rm osc}/(H_{\rm osc}\omega) \ll 1$. This corresponds to our earlier statement that on short time scales (larger energies) there is an approximate discrete symmetry.

An important question is whether we can generalize the symmetry arguments above for the time-dependent functions associated with the non-universal operators in \eqref{Sgpi}-\eqref{Smspi}, \ie $\{m_2,\alpha_i,\beta_i\}$.  On general grounds, in an FRW background described by \eqref{H_osc} we expect that the functions $m_2,\alpha_i,\beta_i$ -- which describe the self-couplings, and couplings/mixings between the Goldstone and the reheat sector $\chi$ -- to be a generic function of the Hubble rate in \eqref{H_osc} and its derivatives. Depending on the couplings between these sectors this suggests that 
in general we can write these functions in the form
\beq\label{G_osc}
F_i(t)= M^p_i(t) P'(\omega' t),
\eeq
where in general the periodic function $P'$ is different from the one in \eqref{H_osc} as is the frequency $\omega'\neq\omega$. Here, the index $i$ collectively represents time-dependent functions $\{m_2,\alpha_i,\beta_i\}$ and $p$ denotes the mass dimension of these functions. Suggested by the symmetry breaking pattern we discussed above, we can similarly take $\dot{M}_i/(M_i \omega) \ll 1$.
\subsubsection{Symmetries of the Action and Implications}

An important consequence of the discrete symmetry of the Goldstone is that non-derivative interactions can appear in the action. When this is a good symmetry we can expand the background and non-universal parameters $\{H,\dot{H},m_2,\alpha_i,\beta_i\}$ in the form
\beq\label{T} 
F_i(t+\pi) = F_i(t) + \dot{F}_i(t)\pi + \fr{1}{2}\ddot{F}_i(t)\pi^2 +\dots  ~.
\eeq 
This breaking is similar in spirit to the work of \cite{Behbahani:2011it}, where those authors considered resonant non-Gaussianity induced through small-scale oscillations in $H$ and $\dot{H}$ during single-field inflation. In the two-sector EFT we are considering here we can extend that study to dynamics that arise in the presence of interactions between the Goldstone $\pi$ and  reheating $\chi$ sectors. 
Moreover, contrary to the situation during inflation, where there is a fixed energy scale corresponding to horizon crossing, \cite{Cheung:2007st}, to study dynamics at the {\it end} of inflation we are often interested in the dynamics at sub-Hubble scales. For sub-Hubble scales with $E_{\rm{probe}} > \omega$ we expect interactions induced by expanding the time-dependent functions in \eqref{T}, which parametrize important contributions to the dynamics. Such interactions can induce large loop corrections for the parameters of the EFT, and additionally back-reaction effects can become large and the perturbative expansion of the EFT of fluctuations will fail. 
In typical studies of preheating, the importance of such contributions correspond to the end of `stage one', which can be followed by turbulence and chaotic behavior \cite{Kofman:1997yn}. We leave an investigation of these stages to future work. In the following, we will focus on the first stages of preheating and establish how our framework captures existing models. We will also explore new models and their connection to observations during the first stages of preheating.   

\subsection{Capturing Existing Models \label{CEM}}
\subsubsection{Reheating Through Self-Resonance}\label{Gsr} 

In this section, we focus on the Goldstone sector in \eqref{Sgpi} to construct models of reheating through self-resonance. 
That is, we want to establish how the EFT reproduces self-resonant models of reheating where inflaton `particles' (here corresponding to the Goldstone $\pi \sim \delta \phi$) are created from oscillations of the background condensate $\phi_{_0}(t)$. We will also consider when gravitational fluctuations can be shown to decouple.  
To begin we expand the time-dependent functions in \eqref{Sgpi} and use the ADM decomposition\footnote{Details appear in Appendix A} of the metric in spatially flat gauge working to second order in fluctuations $\delta N, N^{i}$ and $\pi$. 
We have 
\bea\label{Lcpi}
\nn \mathcal{L}_{\pi_c} &=& \fr{1}{2}\left(\dot{\pi}_c^2-c_\pi^2\fr{(\partial_i \pi_c)^2}{a^2}\right) - \fr{1}{2}m_\pi^2(t) \pi_c^2  
\nn-  \fr{(-2\dot{H})^{1/2}}{c_\pi} \left(\dot{\pi}_c \delta N_c -\fr{1}{2}\left(\fr{\ddot{H}}{\dot{H}}-2\frac{ \dot{c}_\pi}{c_\pi}\right)\pi_c \delta N_c\right)\\
&+& (-2\dot{H})^{1/2} c_\pi ~\left(3H\delta N_c + \partial_i N^{i}_c\right)~ \pi_c + \dots
\eea
where we introduced the canonical fields $\pi_c = \sqrt{-2\dot{H}m_{\rm pl}^2}~c_\pi^{-1} \pi, \, \delta N_c = m_{\rm pl} \delta N, N^{i}_c = m_{\rm pl} N^i$, the sound speed of the fluctuations is
$c_\pi^2 = m_{\rm pl}^2\dot{H}/(m_{\rm pl}^2\dot{H}-m_2^4)$, and we neglect terms involving the scalar curvature as they are sub-leading. 

An important consequence of the background evolution and time-dependent sound speed is that it induces a time-dependent mass\footnote{This is the mass term in the absence of mixing terms given in the second and third lines of \eqref{Lcpi}. } for the Goldstone
\bea\label{mpi}
m_\pi^2 = -3\dot{H}c_\pi^2 - \fr{1}{4}\left(\fr{\ddot{H}}{\dot{H}}-2\frac{ \dot{c}_\pi}{c_\pi}\right)^2 -\fr{3H}{2}\left(\fr{\ddot{H}}{\dot{H}}-2\frac{ \dot{c}_\pi}{c_\pi}\right)-\fr{1}{2} \partial_t\left(\fr{\ddot{H}}{\dot{H}}-2\frac{ \dot{c}_\pi}{c_\pi}\right),
\eea
which we note would vanish in a strictly de Sitter limit with constant sound speed (familiar from the EFT of Inflation).
Resonant effects induced by such time dependence of $c_\pi$ is an interesting possibility that we will explore in future work. For simplicity, here we will focus on the time-dependence of the background and assume that the time dependence of the sound speed is negligible.
 
To understand the Goldstone dynamics we first identify the energy scales at which different phenomena become important. An important scale is the symmetry breaking scale below which we are able to focus on the EFT of the perturbations (we can `integrate out the background') and the Goldstone description can be useful.  Following closely the example of \cite{Baumann:2011su}, we can identify the Noether current associated with the broken symmetry by introducing `fake' Lorentz invariance in \eqref{Lcpi} by rescaling the spatial coordinates 
\beq\label{FLL}
\tilde{\mathcal{L}}_{\rm g} = -\fr{1}{2}(\tilde{\partial}\tilde{\pi}_c)^2 + \dots, 
\eeq  
where $\tilde{x} \equiv c_\pi^{-1} x$, $\tilde{\mathcal{L}}_{\rm g} \equiv c_\pi^3 \mathcal{L}_{\rm g}$ and $\tilde{\pi}_c = (-2\dot{H}m_{\rm pl}^2c_\pi)^{1/2}\pi_c$. The Noether current associated with \eqref{FLL} is then $\tilde{J}^{\mu} = -\Lambda_{\rm sb}^2 \partial^{\mu} \tilde{\pi}_c$, and the symmetry breaking scale is given by\footnote{We present the scale in terms of energy, but it is important to remember that since Lorentz invariance is spontaneously broken energy scales do not necessarily coincide with momenta \cite{Cheung:2007st}.} $\Lambda_{\rm sb}^2 = (-2\dot{H}m_{\rm pl}^2c_\pi)^{1/2}$.

 For the simplest models, with unity sound speed, we have $\Lambda_{\rm sb}^2 = (-2\dot{H}m_{\rm pl}^2)^{1/2}$, and this agrees with expectations that the time evolution of the background is responsible for breaking the time translation symmetry ($H(t)$ is changing in time). In particular, given the background evolution in \eqref{H_osc} we are interested in the time averaged value $\Lambda_{\rm sb}^2 \equiv \langle(-2\dot{H}m_{\rm pl}^2c_\pi)^{1/2}\rangle_T \approx H_\frw m_{\rm pl}~ c_\pi^{1/2}$. For energy scales where $E < \Lambda_{\rm sb}$ the Goldstone description of \eqref{Lcpi} is valid. We emphasize that we are focusing on fluctuations around a decelerating FRW background, and so the symmetry breaking scale is more dependent on time\footnote{This raises the interesting issue of `level crossing', which is ubiquitous when applying EFT to gravitational systems \cite{Burgess:2003jk}.} than the inflationary case \ie $\Lambda_{\rm sb}^2 \propto t^{-1}$. However, in the presence of resonance and with strong enough couplings to the reheating sector  to make reheating efficient,  it is justified to take a decoupling limit $H_\frw \to 0$ and $m_{\rm pl} \to \infty$, such that the combination $H_\frw m_{\rm pl}$ remains fixed. In this case, an evolving symmetry breaking scale is unimportant for the validity of the Goldstone description -- all that is required is a hierarchy of scales $\Lambda_{\rm sb} \gg \omega$ where $\omega$ is the oscillation time scale associated with the background evolution that appeared in \eqref{H_osc}.  

Another important scale in understanding the Goldstone dynamics is the energy scale where mixing with gravitational fluctuations becomes important ($E_{\rm mix})$. Consider the frequency of the Goldstone $\pi_c$ in Fourier space and in the absence of mixing terms 
\beq\label{opi}
\omega_\pi^2 = \fr{c_\pi^2 k^2}{a^2} + m^2_\pi(t) + \dots,
\eeq
where dots represent sub-leading contributions of order $H^2$.  We emphasize that $\omega_\pi$ is the frequency of the Goldstone, whereas the inflaton oscillations have a frequency we continue to denote by $\omega$
which is often comparable to the Goldstone mass $\omega \sim m_\pi$ as follows from \eqref{H_osc} and \eqref{mpi}.  Remembering this distinction, we note that contrary to the inflationary case, we are not interested in the dynamics at a fixed energy scale, and in general whether mixing with gravity is important will depend on the scales one is interested in.  
For example, we can separate the Goldstone modes into relativistic $\omega \lesssim c_\pi k/ a $ (or equivalently $m_\pi \lesssim c_\pi k/ a $) and non-relativistic $\omega>(c_\pi k)/ a$ modes. For relativistic modes, time derivatives scale the same as spatial ones in \eqref{Lcpi}, \ie $\dot{\pi}_c^2\sim c_\pi^2 (\partial_i\pi_c/a)^2\sim \omega_\pi^2 \pi_c$. On the other hand, for non-relativistic modes, spatial derivatives are less important than time derivatives and terms involving the spatial kinetic terms can be compared with the mixing terms in \eqref{Lcpi}. The most relevant mixing term\footnote{Another equally important term is the one proportional to $\dot{\pi}_c \delta N_c$. When we solve for $\delta N_c$ in terms of $\pi_c$ and use this solution in \eqref{Lcpi}, we can integrate by parts the time derivative on $\pi_c$ leading to a term comparable to \eqref{Lmix}.} between $\pi_c$ and gravitational fluctuations is given by
\beq\label{Lmix}
\mathcal{L}_{\rm mix} \supset \fr{(-2\dot{H})^{1/2}}{2c_\pi}\fr{\ddot{H}}{\dot{H}}\pi_c \delta N_c.
\eeq    
From Appendix A, we use the solution $\delta N_c \approx c_\pi \pi_c $ in \eqref{Lmix} and note that $\dot{H}\approx H^2$, $\ddot{H}\approx \omega H^2$ (where we keep the leading terms). This leads to $\mathcal{L}_{\rm mix} \approx \omega H \pi_c^2$ from which we can see the energy scale at which mixing with gravity becomes important is $E_{\rm mix} \approx (\omega H)^{1/2}$. For relativistic modes, mixing with gravity is always irrelevant as $\omega_\pi^2 > \omega^2 \gg \omega H$. For non-relativistic modes, we compare the mixing term with the spatial kinetic term in \eqref{Lcpi}. This leads to the conclusion that mixing with gravity will be important for modes with momenta satisfying the following condition,
\beq\label{kmix}
{\fr{k}{a} \lesssim \fr{\sqrt{\omega H}}{c_\pi}}
\eeq    

\begin{figure}[h!]
\includegraphics[scale=0.99]{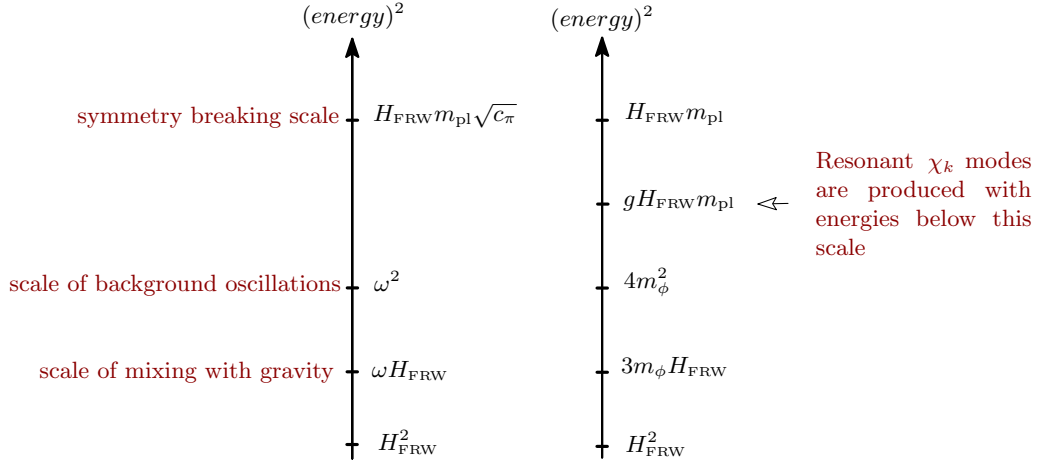}
\caption{Relevant energy scales for the preheating models considered in Section \ref{CEM}. On the left, we have the hierarchy in energy scales associated with the dynamics of the Goldstone boson with a sound speed $c_\pi$ following our general discussion of self-resonant models. The right diagram shows the hierarchy of scales for the example of canonical two-field preheating models.} 
\end{figure}  

{\bf An explicit example:} The generic construction above is useful in studying models of inflaton self-resonance. Consider an example where mixing with gravity at the end of inflation leads to resonant effects for $\pi_c$. For this purpose, we consider a simple limit of the unitary gauge action in \eqref{SU} where $m_2=0$, $m_{\rm pl}^2(3H^2+\dot{H}) = V(\phi_{_0})= m_\phi^2 \phi_{_0}^2/2 ~$, and $\dot{H}m_{\rm pl}^2 = -\dot{\phi}_{_0}^2/2$. These choices correspond to a cosmology dominated by a single scalar field -- the inflaton. In the regime where $m_\phi \gg H$, the background condensate oscillates around the minimum of its potential $V = m_\phi^2 \phi_{_0}^2/2$, and in this case we can solve for the background evolution \cite{Mukhanov:2005sc}
\bea \label{ExB}
H(t)&=&H_{\frw}(t) - \frac{3H_\frw(t)^2}{4m_\phi}\sin (2m_\phi t) + \ldots, 
\eea
where $H_\frw = 2/(3t)$ is the Hubble rate in a matter dominated universe with scale factor $a(t) \propto t^{2/3}$ and dots represent terms suppressed by higher powers of $H_m/m_\phi$. This solution has exactly the form proposed in \eqref{H_osc} with $H_{\osc} \equiv  -3H_\frw^2/4m_\phi$, $\omega \equiv 2m_\phi$, and we also have $H_\frw \ll m_\phi$.

Given the background evolution in \eqref{ExB}, we can now consider the dynamics of $\pi_c$. To reproduce this class of models we take the $c_\pi \to 1$ limit, and solve for the constraints $\delta N_c$ and $N^{i}_c$. Using our results from Appendix A, along with \eqref{Lcpi} we have
\beq\label{ExLpi}
\mathcal{L}_{\pi_c} = -\fr{1}{2}(\partial\pi_c)^2 - \fr{1}{2}\Big(m_\pi^2(t) + m_{\rm mix}^2(t)\Big) \pi_c^2,  
\eeq
where the mass mixing induced by $\delta N_c$ and $N^{i}_c$ is 
\beq\label{Exmmix}
m_{\rm mix}^2 = 6\dot{H} + 2\fr{\ddot{H}}{H}-2\fr{\dot{H}^2}{H^2}.
\eeq
Using the background evolution given by \eqref{ExB} and \eqref{Exmmix} the mode equation for the re-scaled field variable $\tilde{\pi}_c = a^{3/2}\pi_c$ can be written as 
\beq\label{meqpi}
\ddot{\tilde{\pi}}_c + \left[   \fr{k^2}{a^2}+m_\phi^2\left(      1+6\fr{H_\frw}{m_\phi}\sin(2m_\phi t)    \right) \right]\tilde{\pi}_c = 0,
\eeq
where we have dropped additional terms further suppressed by $H_\frw^2/m_\phi^2$ and $m_\pi^2 \to V''(\phi_{_0})= m_\phi^2$ which follows from relating derivatives of the potential to the time derivatives of the Hubble rate given in \eqref{ExB} (See Appendix B). 

To establish whether self-resonance results in particle production we can recast \eqref{meqpi} in the form of a Mathieu equation by re-defining the time variable $z=m_\phi t +\pi/4$ with $A_k = 1 + k^2/(a^2m_\phi^2)$ and $q =3H_\frw/m_\phi$. As the background evolution implies the hierarchy $H_\frw \ll m_\phi$, this implies modes in equation \eqref{meqpi} will be in the narrow resonance regime, $q \ll 1$. The first instability corresponds to the condition $A_k < 1 + q$ implying modes with momenta 
\beq
\fr{k}{a} < \sqrt{3H_\frw m_\phi}
\eeq
will be amplified \cite{Kofman:1997yn}. This result matches well with our previous estimate on the momentum scales where mixing with gravitational fluctuations is important in \eqref{kmix} (recalling we have $c_\pi=1$ here). 

Such resonant effects due to mixing with gravity have been considered previously in the literature \cite{Easther:2010mr,Jedamzik:2010hq}, 
where those authors studied the growth of the density perturbations and the onset of non-linear effects arising during oscillations of the background. 
Here, we can use the EFT to reproduce their results 
\beq
\delta_k \equiv \fr{\delta \rho_k}{\bar{\rho}(t)} = \fr{\delta \rho_k}{3H^2m_{\rm pl}^2} \propto \left(\fr{k}{aH_\frw}\right)^2, ~~ {\rm for} ~~~ \fr{k}{a}<\sqrt{3H_\frw m_\phi},
\eeq 
where $\delta \rho_k$  is defined as 
\beq
\delta \rho_k = (-2\dot{H})^{1/2}m_{\rm pl} \left[\dot{\pi}_c-\fr{1}{2}\left(3+\fr{\ddot{H}}{\dot{H}}-2\fr{\dot{H}}{H}\right)\pi_c\right].
\eeq

We now consider how the EFT captures models where the reheat sector results from the inflaton resonance given by the time-dependent functions in \eqref{Sspi} and \eqref{Smspi}. 
If any of these couplings are stronger than gravitational strength the resonance in the reheat sector will typically dominate over the gravitationally induced effects discussed above.  

\subsubsection{Reheating in a Two-Field Model \label{RRS}}
In this section, we explicitly demonstrate how the EFT approach reproduces models of two-field
reheating, taking as a concrete example the specific class of models given by \eqref{toyL}. In the early
stages of preheating the inflaton will dominate the energy density. We take the reheat field to be
initially in its vacuum\footnote{We saw in Section \ref{sec2} that it was a challenge for the background EFT model, but this is natural here as the shift
symmetry of the background has been badly broken by the interactions.} with $\chi_{_0} = 0$, and we consider production of $\chi$ quanta in the presence of the
oscillating inflaton condensate $\phi_{_0}(t)$. In the unitary gauge with $\phi = \phi_{_0}(t)$ and $\chi_{_0} = 0$, we have the following matter
Lagrangian
\beq
S_{m} = \int d^4 x\sqrt{-g}\left[-\fr{1}{2}\dot{\phi}_{_0}^2 g^{00}-V(\phi_{_0})-\fr{1}{2}g^{\mu\nu}\partial_{\mu}\chi\partial_{\nu}\chi-\fr{1}{2}(U''(\chi_{_0})+g^2\phi_{_0}^2)\chi^2\right].
\eeq
Using the background equations of motion we can cancel the tadpole terms, $m_{\rm pl}^2(3H^2+\dot{H}) = V(\phi_{_0})= m_\phi^2 \phi_{_0}^2/2 ~$, $\dot{H}m_{\rm pl}^2 = -\dot{\phi}_{_0}^2/2$, and the unitary gauge matter Lagrangian is then given by
\beq\label{Lmc}
\mathcal{L}_{m} =m_{\rm pl}^2 \dot{H} g^{00}-m_{\rm pl}^2(3H^2+\dot{H})-\fr{1}{2}g^{\mu\nu}\partial_{\mu}\chi\partial_{\nu}\chi-\fr{1}{2}\left(m_\chi^2+2 \fr{g^2m_{\rm pl}^2}{m_\phi^2}(3H^2+\dot{H})\right)\chi^2,
\eeq
where we defined $U''(\chi_{_0})\equiv m_\chi^2$.
Comparing with the unitary gauge action \eqref{SU} -- \eqref{Sm}, the matter Lagrangian \eqref{Lmc} corresponds to the following choice for non-universal parameters in the EFT framework, 
\beq\label{ceXp}
\alpha_1 = 1,~~~~ \alpha_3=m_\chi^2 + 2 \fr{g^2m_{\rm pl}^2}{m_\phi^2}(3H^2+\dot{H}), ~~~ \{m_2,\alpha_2,\alpha_4,\beta_1,\beta_2\}=0.
\eeq

We emphasize that in this model the linear mixing between the $\chi$ sector and gravitational sector (which includes the Goldstone in the unitary gauge) vanishes automatically since $\beta_1,\beta_2 = 0$ in \eqref{Sm}. As before, we can introduce the Goldstone sector in \eqref{Lmc} following the transformation\footnote{It is important to note that the transformation $t \to t + \pi$ that introduces the Goldstone also induces non-linear interactions between the Goldstone and reheat sectors -- we will elaborate on this below.} rules in \eqref{transform}.  However, in the presence of strong resonance in the $\chi$ sector, \ie if $\dot{\alpha_3}/\alpha_3^2 > \mathcal{O}(1)$ during any time in the linear stage of preheating, Goldstone fluctuations will be negligible compared to the $\chi$'s that are amplified through the strong resonance. In general, the validity of this argument relies on 
the strength of the coupling between the background and the $\chi$ sector through the mass term. For example, in the model we are considering here, introducing $\pi$ via $t \to t+\pi$ (See also \eqref{transform}) will lead to the Goldstone sector we have discussed in the previous section, where mixing with gravity leads to weak resonance $q \approx H_\frw / m_\phi \ll 1$ ({\it c.f.} \eqref{meqpi} and the discussion that follows). On the other hand, the strength of the resonance in the $\chi$ sector depends on the ratio $g m_{\rm pl} /m_\phi$ which can be quite large unless $g \ll 1$. Too see this in detail, it is enough to compare the scales in our EFT. The strength of the resonance in $\chi$ can be read from \eqref{ceXp} and compared to the strength $\approx m_\phi H_\frw$ of the resonance in the Goldstone sector in equation \eqref{meqpi}.  The following condition is sufficient to neglect the Goldstone dynamics
\beq
 g^2 \left(\fr{m_{\rm pl}}{m_\phi}\right)\left(\fr{\Lambda_{\rm sb}}{m_\phi}\right)^2 > 1 . 
\eeq
It is clear from this expression that unless the coupling constant is tiny $g  \ll 1$ we can neglect the mild amplification of Goldstone due to mixing with gravity. 

Another simplification we can make in this case is to consider the decoupling limit in the EFT where $|\dot{H}|\approx H_\frw^2 \to 0$ and $m_{\rm pl}^2 \to \infty$, while keeping the combinations $\dot{H}m_{\rm pl}^2$ and $H^2 m_{\rm pl}^2$ as constant. In this limit, it is clear that $\pi$ fluctuations will stay in their vacuum as the terms leading to narrow resonance vanishes ($ H_\frw \to 0$). We also note that the decoupling limit corresponds to taking the rigid space-time limit, $a \to 1$ that is commonly discussed in the preheating literature\footnote{An additional and important point on the decoupling limit is that in this limit the time-dependent functions such as $\alpha_3$ we are considering will be purely periodic functions. This can be seen by using \eqref{ExB} in equation \eqref{ceXp} and taking the decoupling limit. This implies that EFT should respect an exact discrete symmetry in this limit.} \cite{Kofman:1997yn,Amin:2014eta}.

To study particle production, we can focus on the decoupling limit of the Lagrangian \eqref{Lmc}, and consider the mode equation for $\chi$ as,  
\beq
\ddot{\chi}_k + \omega_\chi^2(t)\chi_k = 0
\eeq  
where the time dependent frequency is given by
\beq
\omega_\chi^2 = k^2 + m_\chi^2 + \fr{g^2m_{\rm pl}^2}{m_\phi^2}(3H^2+\dot{H}).
\eeq 
In the decoupling limit, the time dependent mass induced by the background evolution stays intact, which is crucial for particle production. As we have mentioned before, particle production corresponds to the breakdown of the adiabaticity in the frequency, \ie $|\dot{\omega}_\chi/\omega_\chi^2| > \mathcal{O}(1)$. Using \eqref{ExB} and the relations with the potential and Hubble rate in Appendix B, this condition translates into
\beq
K^2 \lesssim g ~H_\frw m_{\rm pl} \approx g \Lambda_{\rm sb}^2,
\eeq
where $K=\sqrt{k^2 + m_\chi^2}$ is the rescaled momenta. In the example we are considering, we see that this condition justifies the use of the EFT formalism as the resonant modes have a momenta
much smaller than the symmetry breaking scale for small enough coupling, \ie $H_\frw m_{\rm pl} \equiv \Lambda_{\rm sb}^2 \gg g \Lambda_{\rm sb}^2 $ for $g \ll 1$. The structure of the instability band along with the exponentially growing solutions in the $\chi$ sector have been studied many times in the literature \cite{Amin:2014eta}. Here, our main purpose is to show the connection of the EFT approach to well established two-field reheating models. 

Another potential use of EFT formalism is to capture the effects of backreaction. This can be achieved by realizing that once we introduce the Goldstone mode in the unitary gauge Lagrangian \eqref{Lmc} the time dependent mass (and for general models other time dependent functions) of $\chi$ becomes $\alpha_3(t+\pi)$. As $\alpha_3$ is a rapidly varying function of time in the presence of particle production in the $\chi$ sector, this term will induce higher order interactions between $\pi$ and $\chi$ upon expanding the function, 
\beq\label{Lexi}
\mathcal{L}_{int} = -\fr{1}{2}\left(\dot{\alpha}_3\pi + \fr{1}{2}\ddot{\alpha}_3\pi^2\right)\chi^2.
\eeq     
In particular, in the current example the first term in \eqref{Lexi} will lead to a tadpole term for $\pi_c = (-2\dot{H})^{1/2}m_{\rm pl} \pi$. In the Hartree approximation \cite{Kofman:1997yn} this gives  
\beq
\mathcal{L}_{int} \supset -\fr{1}{2}\fr{\dot{\alpha}_3}{(-2\dot{H}m_{\rm pl}^2)^{1/2}}\langle\chi^2\rangle\pi_c,
\eeq 
where 
\beq
\langle\chi^2(t)\rangle = \fr{1}{2\pi^2}\int_0^{\infty} dk~ k^2~ |\chi_k(t)|^2.
\eeq
The existence of such a tadpole term can be considered as an indication of backreaction effects. For example, as we produce $\chi$ particles the coefficient in front of $\pi_c$ will grow and may eventually disturb the background evolution. In particular they can increase the frequency of the background oscillations of the condensate \cite{Kofman:1997yn},
\beq\label{Cm}
m_\phi^2 \to m_\phi^2 + \fr{\dot{\alpha}_3}{(-2\dot{H}m_{\rm pl}^2)^{1/2}}\langle\chi^2\rangle
\eeq
In order to understand the onset of the backreaction effects in the presence of particle production, we can compare the second term in \eqref{Cm} with $m_\phi^2$. We refer to this time where the backreaction becomes important as $t_b$ and the condition reads  
\beq
m_\phi^2 = \fr{\dot{\alpha}_3(t_b)}{(-2\dot{H}(t_b)m_{\rm pl}^2)^{1/2}}\langle\chi^2(t_b)\rangle
\eeq
Knowing the solutions for $\chi_k$, the background evolution \eqref{ExB} and the couplings $\alpha_3$ one can calculate $t_b$.
 
We emphasize that our discussion in this section is not limited to the example given by \eqref{ceXp}. Using the EFT formalism, we can in principle capture models that belong to the same ``universality class'', \ie direct coupling models with interactions including $\mathcal{L}_{m} \propto \mu \phi \chi^2 $ and non-renormalizable couplings $\mathcal{L}_{m}\propto \phi^{n}\chi^2/M^{n-2}$ where $n>2$ and $M, \mu$ are energy scales \cite{Dufaux:2006ee}.    

\subsection{A New Class of Reheating Models \label{new}}
In the previous section, we showed how the EFT captures resonance effects in two-field reheating models. 
We now reconsider particle production in the presence of a reduced sound speed for the reheat field, $c_\chi \neq 1$.  
Familiar from the EFT of Inflation and Dark Energy, there is no symmetry protecting $c_\chi =1$ in the EFT of reheating.
This gives rise to a new class of models for preheating where the produced particles can have $c_\chi \ll 1$.

We follow our previous discussion in Section \ref{BevS} and consider the time-dependent functions associated with the reheat sector $\{\alpha_i,\beta_i\}$.
The terms proportional to $\beta_1$ and $\beta_2$ in \eqref{Smspi} lead to mixing of $\chi$ with both gravity and the Goldstone sector.  We will ignore these terms here, leaving a discussion of them to Appendix A. In the absence of these mixing terms we focus on the action \eqref{Sspi}. Defining the canonical field $\chi_c = \alpha_\chi(t) \chi$ where $\alpha_\chi^2(t) = \alpha_1(t) + \alpha_2(t) $, we have the following second order Lagrangian for the canonical reheat field 
\beq
\mathcal{L}_{\chi_c} = \fr{1}{2}\left[\dot{\chi}_c^2 - c_\chi^2(t) \fr{(\partial_i\chi_c)^2}{a^2}\right] - \fr{1}{2} m_\chi^2(t)\chi_c^2,
\eeq
where we have defined the sound speed $c_\chi^2 = \alpha_1 / (\alpha_1 + \alpha_2)$ and the time-dependent mass term is 
\beq\label{msgm}
m_\chi^2(t)= \fr{\alpha_3(t)}{\alpha_\chi^2(t)} - {\left(\frac{\dot{\alpha_\chi}}{\alpha_\chi}\right)^2} + 3H {\left(\frac{\dot{\alpha_\chi}}{\alpha_\chi}\right)} + \partial_t{\left(\frac{\dot{\alpha_\chi}}{\alpha_\chi}\right)}.
\eeq
Similar to the Goldstone case in Section \ref{Gsr}, we have a time-dependent mass $m_\chi(t)$ induced by the time dependence of the sound speed $c_\chi$ and $\alpha_1$ \footnote{Recall that $c_\chi^2 = \alpha_1 /\alpha_\chi^2$}. We will concentrate on strong resonant effects due to non-adiabaticity in the time-dependent coefficient $\alpha_3$ and assume that the time variation of $\alpha_\chi$ is slow compared to $\alpha_3$, so that the sound speed is nearly constant\footnote{Again, we leave the interesting case of strong time dependence of the sound speed to future work.} (where $\alpha_1, \alpha_2 \approx constant$). We can then neglect the 
last three terms in \eqref{msgm} and the mode equation for the re-scaled field variable $\tilde{\chi}_c = a^{3/2} \chi_c$ in Fourier space is
\beq
\ddot{\tilde{\chi}}^{k}_c + \left[c_\chi^2  \frac{k^2}{a^2} +\alpha_3+\Delta\right]\tilde{\chi}^{k}_c = 0,
\eeq  
where $\Delta = -3(3H^2+2\dot{H})/4 \approx \mathcal{O}(H^2)$ are gravitational terms resulting from the rescaling $\chi_c \to \tilde{\chi}_c$ and we have absorbed the constants $\alpha_1, \alpha_2$ into the definition of $\alpha_3$. Following our discussion in Section \ref{BevS}, it is convenient to parameterize $\alpha_3$ as $\alpha_3=M^2(t) F(\omega t)$, where $M(t)$ is always adiabatic so that $\dot{M}/M^2 \ll 1$ and $F$ is a periodic function which must violate adiabaticity so that preheating occurs. That is, at some point adequate particle production requires the so-far arbitrary function to satisfy $\dot{F}/F^2 >1$. In many models the periodicity of the function will be set by the background evolution in \eqref{H_osc}. We focus on the strong resonance regime where $M \gg H$ and $M/\omega \gg 1$ and hence drop $\mathcal{O}(H^2)$ terms in the frequency $\omega_\chi^2$, 
\beq
\omega_\chi^2 = c_\chi^2 \frac{ k^2}{a^2} + M^2 F(\omega t).
\eeq
The non-adiabaticity in $\alpha_3$ will lead to non-adiabaticity in the frequency $\omega_\chi^2$, \ie $\dot{\omega}_\chi/\omega_\chi^2 > \mathcal{O}(1)$.  We take this to occur as times $t_j$ when $\omega_\chi^2$ is at its minimum\footnote{Note that here we are focusing on non-tachyonic resonance, for tachyonic resonance this situation will be different, see \eg \cite{Dufaux:2006ee}.}. This suggests that we can expand the frequency around the times $t_j$ as 
\beq
\omega^2_\chi \simeq c_\chi^2 \fr{ k^2}{a^2} + \fr{1}{2} M^2 \omega^2 (t-t_j)^2 + \dots
\eeq
where we have used $\ddot{F}\approx \omega^2 F$ and dots represent higher order terms in the $t-t_j$ expansion. This allows us to re-write the mode equation in a simpler form
\beq\label{meqs}
\ddot{\tilde{\chi}}^{k}_c + \left[c_\chi^2 \frac{ k^2}{a^2} +\fr{M^2\omega^2}{2}(t-t_j)^2\right]\tilde{\chi}^{k}_c = 0
\eeq
and the typical momenta when adiabaticity is violated $\dot{\omega}_\chi > \omega^2_\chi$ corresponds to
\be\label{Rm}
k_\ast^2 \equiv \frac{M\omega}{c_\chi^2} \gtrsim \frac{k^2}{a^2},
\eeq
We see that for $c_\chi < 1$, the physical wave numbers inside the resonant regime are further enhanced (the resonance band is broadened) compared to the standard cases that have been studied in the literature. It is customary to map the mode equation \eqref{meqs} to a scattering problem described by a Schr\"{o}dinger equation with a negative parabolic potential by defining a new time variable $\tau \equiv c_\chi k_* (t-t_j)$ and a dimensionless physical momentum $\kappa \equiv k/(ak_*)$,
\beq\label{Sseq}
\fr{d^2 \tilde{\chi}^k_c}{d\tau^2} + \left(\kappa^2 + \tau^2\right)\tilde{\chi}^k_c = 0.
\eeq
The solution to the scattering problem and the resulting number density of particles between scattering events has appeared in the literature many times \cite{Kofman:1997yn,Bassett:2005xm} (See also \cite{Amin:2015ftc}). In real space, the growth of the number density of particles can be described by the following expression \cite{Kofman:1997yn},
\beq\label{ns}
n_\chi(t) = \fr{1}{2\pi a^3}\int d^3 k ~n^k_\chi(t) \sim \fr{k_{*}^3}{\sqrt{\pi \mu m_\phi t}} e^{2\mu m_\phi t},
\eeq
where (for simplicity) we have assumed that the background is given by the quadratic potential we considered before, \ie $\omega \sim  m_\phi$. Here $\mu$ is the maximum value of the Floquet index at $k_{\max} \approx k_* /2$ \cite{Kofman:1997yn}. It is clear from this expression that there will be an enhancement in the number of produced particles due to the small sound speed in the $\chi$ sector, $k_* \propto c_\chi ^{-1}$. This also agrees with our intuition as equation \eqref{Rm} tells us that resonant bands are wider for $c_\chi < 1$ and thus the contribution to the integral in \eqref{ns} over resonant modes will be enhanced by factors of $c_\chi^{-1}$. In the next section we will consider observational consequences of the EFT of reheating, focusing on this new class of models with non-standard sound speed.  We also discuss additional challenges and future directions for the approach.

\section{Challenges and Outlook\label{sec4}}
In this paper we have presented an EFT approach to reheating that overcomes the challenges of the background evolution discussed in Section \ref{sec2} and is adequate to capture all existing reheating models in the literature.
Guided by symmetries, our approach is also useful for finding new models of reheating, \eg we found a new class of models where the reheating sector has $c_\chi \neq 1$.  
However, there are many challenges remaining for our EFT approach. 

One of the more serious concerns is the lack of a direct connection to observations. 
This problem is not specific to our approach, with the lack of direct observational constraints on reheating being an important reason that far less is known about this epoch than inflation.
In our EFT framework, symmetries help to alleviate more of the theoretical uncertainties associated with reheating than a toy model approach.  For example, the need to non-linearly realize time translations demonstrated that many of the unknown coefficients are related, and the need to violate non-adiabaticity (required for particle production) also placed some level of theoretical constraint on the reheating sector. 
Nevertheless, we saw in Section \ref{sec3} there are a large number of free functions that must be further restricted by observations.  Unlike the situation for inflation, where non-Gaussianity and features in the primordial power spectrum are a rich source of observational constraints, direct observational constraints on reheating are lacking.  One possibility to remedy this is gravitation wave (GW) signatures. 

Once particles are produced during reheating\footnote{This should not be confused with sourcing a gravity perturbation with a second order scalar perturbation. Here we are considering on-shell particles that are classically scattering off of each other and generating a GW spectrum.  We refer the reader to \cite{Dufaux:2007pt} for more details.} they can scatter off each other creating a background of GWs \cite{Easther:2006gt,Easther:2006vd}.  The scattering leads to a transverse-traceless source for the gravitons
\beq\label{eqh}
\ddot{h}_{ij} + 3H\dot{h}_{ij} - \fr{1}{a^2}\partial^2 h_{ij} = \fr{2}{m_{\rm pl}^2} T^{TT}_{ij}
\eeq
Following the methods of \cite{Dufaux:2007pt} we can then estimate the critical density of gravitational waves today\footnote{For a different approach we refer the reader to \cite{Giblin:2014gra}.}
\beq\label{Ogw}
\Omega_{\rm gw} = \fr{S_k(t_f)}{a_J^4~ \rho_J}~ \left(\fr{a_J}{a_{\rm rh}}\right)^{1-3w}\left(\fr{g_{\rm rh}}{g_0}\right)^{-1/3} \Omega_{r,0},
\eeq
where subscript ``$0$'' denotes a quantity evaluated today, `$ J $ ' represents the time when the universe becomes radiation dominated and `${\rm rh}$' denotes the beginning of reheating. Here, $\omega$ is the average equation of state of the universe between the time interval $t_J<t<t_{\rm rh}$ and $g_i$ is the effective relativistic degrees of freedom.  Finally, the source term
$S_k$ encodes the predictions for different classes of models in the EFT. 

For example, let us consider the new class of models discussed in Section \ref{new}. In that case the source term $S_k$ is given by
\bea\label{Sk}
S_k(t_f) &=& \fr{c_\chi^4~ k^3}{4\pi^2 m_{\rm pl}^2}\int dp \int_{-1}^{1} d(\cos\theta)~ p^6 \sin^4\theta  \\
\nn &\times& \Bigg[\left\vert\int_{t_i}^{t_f} dt \cos\left(kt\right)\chi_c(p,t)\chi_c(|\vec{k}-\vec{p}|,t)\right\vert^2 +\left\vert\int_{t_i}^{t_f} dt \sin\left(kt\right)\chi_c(p,t)\chi_c(|\vec{k}-\vec{p}|,t)\right\vert^2\Bigg]
\eea 
where we focus on two-body scattering, $\theta$ is the scattering angle, and we assume that scattering happens at a fast enough rate that we can neglect the Hubble expansion. 
To get an order of magnitude estimate we can focus on the low momenta.  In this case, the contribution of the mode functions to time integrals will be maximal for $p_* = \sqrt{M\omega}/c_\chi$ 
and defining a dimensionless momentum $P = p/p_*$ we have
\beq\label{Sk2}
S_k^{j+1} \sim \fr{1}{c_\chi^3} \fr{(M\omega)^{3/2} k^3}{ m_{\rm pl}^2} \int_{-1}^{1}d(\cos\theta) \sin^4\theta \int ~ dP P^6 \times [Time~integrals],
\eeq 
where we recall that $\alpha_3$ is parameterized by $M$ and $\omega$ as in \eqref{meqs}, and so the EFT parameters are determining the strength of the GW signal.
Moreover, the gravitational waves will be amplified by a factor of $c_\chi^{-3}$. This scaling may be counter-intuitive to the reader.  The prefactor in \eqref{Sk} results from the two-to-two scattering of the particles as their momenta is now $p \rightarrow c_\chi p$. However, the lower sound speed implies it costs less energy to produce the particles leading to an enhancement of the particle production rate, and more particles scattering leads to more gravity waves.  Thus, the GW signal is enhanced compared to the $c_\chi=1$ case. 
Assuming this signal survives the later stages of reheating the detectability will depend on the peak frequency \cite{Easther:2006gt,Easther:2006vd,Giblin:2014gra}
\beq
f = \frac{\sqrt{M \omega}}{a_j \rho_j^{1/4} c_\chi} \; 4 \times 10^{10} \; \mbox{Hz},
\eeq
which again depends explicitly on the EFT parameters and the sound speed.
We see that by reducing the sound speed we can increase the frequency in the new class of reheating models. 

GWs provide one way to constrain the EFT parameters. However, we leave a more complete analysis, which requires following the signal through all the stages of reheating\footnote{One interesting approach would be to 
see if we could combine the EFT framework here with the recent fitting analysis of \cite{Figueroa:2016wxr}.}, to future work.
Primordial Black Hole constraints and the matching of inflationary perturbations to late time observables lead to additional ways in which the EFT parameters may be restricted.  
In regards to the latter, we have stressed that direct observables correspond to perturbations, however the subtle ways in which we match inflationary predictions to CMB and LSS observations does depend implicitly on the background dynamics, particularly through the equation of state. Recently, it has been shown that the physics of reheating (including non-linearities and back-reaction) can have subtle and interesting effects on the equation of state 
and the dynamics of thermalization \cite{Lozanov:2016hid}.  We hope to return to these issues and interesting possibilities in future work.

In addition to the challenge to connect with observations, a number of theoretical issues remain to be addressed.
In particular, in this paper we have primarily focused on connecting the EFT to scalar field driven models of reheating.
However, the spectator field $\chi$ can be thought of as an additional clock field, which can also represent reheat fields beyond spin zero.
Extending our framework to other spins is an important consideration.  We have also primarily focused on the first stage of reheating in the EFT.
However, one of the most useful applications of our approach could be to gain a better understanding of the rescattering and back-reaction effects that happen following the first stage.
These are stages that usually require lattice simulations, and the Goldstone approach could be a fruitful way to get a better analytical understanding.
There is also the issue of when the produced particles become significant enough that they contribute to the energy density.  At this point the Goldstone boson (related to the matter sector responsible for time-translations being broken) can change its nature from inflatons to the reheat field. How this transition proceeds is important for establishing the connection between the Goldstone and the background fields.  This is similar to the situation 
in studies of dissipation in the EFT of Inflation (see \eg \cite{LopezNacir:2011kk}), and we expect many of the techniques there could prove useful for the case of reheating as well.

\section*{Acknowledgements}
We thank Peter Adshead, Rouzbeh Allahverdi, Mustafa Amin, Daniel Baumann, Robert Brandenberger, Sera Cremonini, Jay Hubisz, Shinji Mukohyama, 
Jayanth Neelakanta, Sonia Paban, Marco Peloso, and Jun'ichi Yokoyama for useful discussions.  We are especially grateful to Masahide Yamaguchi for many useful conversations and hospitality while this work was completed, and to Kuver Sinha for initial collaboration on this project.  OO would like to thank JP van der Schaar and the University of Amsterdam for hospitality. 
This work was supported in part by NASA Astrophysics Theory Grant NNH12ZDA001N, DOE grant DE-FG02-85ER40237, National Science Foundation Grant No. PHYS-1066293 and the hospitality of the Aspen Center for Physics. 

\begin{center}
This paper is dedicated to the memory of Lev Kofman.
\end{center}
\newpage

\section*{Appendix A: ADM Formalism and Mixing with Gravity}
To account for gravitational fluctuations and discuss the  regime where they are irrelevant to the dynamics of the Goldstone we decompose the metric in the ADM form. In the spatially flat gauge we have
\begin{equation}
ds^2=-(N^2-N_iN^i)dt^2+2N_i dx^i dt+\hat{g}_{ij}\,dx^idx^j,
\end{equation}
where $\hat{g}_{ij} = a^2 (\delta_{ij}+h_{ij})$ is the spatial metric and our gauge choice implies $h_{ii}=\partial_i h_{ij}=0$. Inverse metric elements can be written as 
\begin{equation}
g^{00}=-\frac{1}{N^2}\,,
\quad
g^{0i}=g^{i0}=\frac{N^i}{N^2}\,,
\quad
g^{ij}=h^{ij}-\frac{N^iN^j}{N^2}\,.
\end{equation}

To find the relevant terms in the gravitational sector, we expand the Einstein Hilbert term as
\beq
\label{EH} S_{\rm g} \supset \fr{m_{\rm pl}^2}{2} \int \! d^4 x \:
\sqrt{-g} \, R = \fr{m_{\rm pl}^2}{2} \int \! d^4 x \:
\sqrt{\hat g} \, \big[ N R^{(3)} + \frac{1}{N} (E^{ij} E_{ij} -
E^i{}_i {}^2) \big],
\eeq   
where $R^{(3)}$ is the three curvature associated with spatial metric $\hat{g}_{ij}$ and $E_{ij}$ is related to the extrinsic curvature of constant time slices through
\beq\label{extrinsic}
E_{ij} \equiv N K_{ij} = \fr{1}{2} [{\partial_t
{\hat g}}_{ij} - \hat \nabla_i N_j - \hat \nabla_j N_i] \; ,
\eeq 
where $\hat{\nabla}_i$ is the covariant derivative with respect to spatial metric $\hat{g}_{ij}$. Using the above expressions, we can expand \eqref{Sgpi} up to second order in scalar fluctuations 
\bea\label{Smix0}
\nn
S_{\rm g}=\int d^4x \,a^3\Bigg[&-&\frac{m_{\rm pl}^2\dot{H}}{c_\pi^{2}}\Big(\dot{\pi}^2-c_\pi^2\frac{(\partial_i\pi)^2}{a^2}\Big)-3m_{\rm pl}^2\dot{H}^2\pi^2 +m_{\rm pl}^2(2c_\pi^{-2}\dot{H}\dot{\pi}-6H\dot{H}\pi)\delta N
+2m_{\rm pl}^2\dot{H}N^i\partial_i\pi
\\
&-&
m_{\rm pl}^2(3H^2+c_\pi^{-2}\dot{H})\delta N^2
-2m_{\rm pl}^2H\delta N\partial_i N^i
\Bigg]\,
\eea
where the speed of sound is defined as $c_\pi^2 = m_{\rm pl}^2\dot{H}/(m_{\rm pl}^2\dot{H}-m_2^4)$. Defining the canonical fields, $\pi_c = \sqrt{-2\dot{H}m_{\rm pl}^2}~c_\pi^{-1} \pi, \delta N_c = m_{\rm pl} \delta N, N^{i}_c = m_{\rm pl} N^i$, one can re-write the Lagrangian as in \eqref{Lcpi}. 

Focusing on the Goldstone sector for now, we can solve for the Lagrange multipliers $\delta N$ and $N^{i}$ in terms of $\pi$. To linear order in $\pi$ we have, 
\beq\label{adm2}
\delta N = -\fr{\dot{H}}{H}\pi, ~~~~~~~ \partial_i N^{i} = c_\pi^{-2}\fr{\dot{H}}{H^2}\partial_t \left(H\pi\right).
\eeq  
Using the canonical field definitions above we may write
\beq
\delta N_c = \fr{(-2\dot{H})^{1/2}}{2H}\pi_c, ~~~~~~~ \partial_i N^{i}_c = c_\pi^{-2}\fr{\dot{H}}{H^2}\partial_t \left(\fr{c_\pi H\pi_c}{(-2\dot{H})^{1/2}}\right).
\eeq
Using these solutions for the gravitational fluctuations $\delta N_c$ $N^{i}_c$ in \eqref{Lcpi} (while taking the $c_\pi \to 1$ limit) we recover the result of \eqref{ExLpi}.

In the presence of a reheat sector $\chi$, we need to take into account the mixing between $\chi$ and gravitational fluctuations, as well as $\pi-\chi$ mixings. Considering the mixings at second order we need to take into account the action in \eqref{Smspi}. Expanding up to second order in $\delta N$, $N^{i}$, $\pi$ and $\chi$, we have
\beq\label{Smix}
S_{\rm mix}^{(2)}=\int d^4 x\,a^3\Big[
2\beta_1\left(\delta N-\dot{\pi}\right)\chi-2\beta_2\left(\delta N-\dot{\pi}\right)\dot{\chi}
\Big]\, .
\eeq
We note that the action \eqref{Sspi} does not lead to any second order mixing therefore it is enough to consider the mixing action above. Combining \eqref{Smix0} and \eqref{Smix} in the presence of mixing we have the following solutions for the constraints, 
\beq
\delta N=-\frac{\dot{H}}{H}\pi, ~~~~~ \partial_i N^{i} = c_\pi^{-2}\fr{\dot{H}}{H^2}\partial_t \left(H\pi\right) +\frac{\beta_1}{m_{\rm pl}^2 H}\chi
-\frac{\beta_2}{m_{\rm pl}^2 H}\dot{\chi} .
\eeq
We see that inclusion of reheat sector does not change the solution for $\delta N$, but we have additional contributions to $N^{i}$ proportional to the time-dependent parameters $\beta_1, \beta_2$ . To illustrate the decoupling of $\chi$, we consider a simple $\pi_c$ sector with $c_\pi =1$ and note that time derivatives of canonically normalized fields $\chi_c$ and $\pi_c$ have the approximate scalings in the WKB approximation,
\beq
\dot{\pi}_c \approx \omega_\pi \pi_c \sim \omega  \pi_c, ~~~~ \dot{\chi}_c \approx \omega_\chi \chi_c \sim \sqrt{\alpha_3} \chi_c \sim M \chi_c,
\eeq
where we take $|\alpha_3| = M^2$ following our discussion in the main text and focused on the non-relativistic modes for both fields. Following our discussion in section \ref{BevS}, we assume that the strength of the couplings $\beta_1$ and $\beta_2$ is as strong as the time-dependent parameter $\alpha_3$ responsible for the resonance. By dimensional analysis, we therefore take $|\beta_1| \sim M^3 $ and $|\beta_2|\sim M^2$. Canonically normalizing the fields as before we find from \eqref{Smix} that for resonant modes mixing between $\chi_c$ and gravitational fluctuations can be neglected in the following range of momenta
\beq
\left(\fr{M}{\Lambda_{\rm sb}}\right) \sqrt{MH} < \fr{c_\chi k}{a} < \sqrt{M \omega} .
\eeq 
Similarly we have the following range where we can neglect direct mixing between $\pi_c$ and $\chi_c$,
\beq
\left(\fr{M}{\Lambda_{\rm sb}}\right) \sqrt{M\omega} < \fr{c_\chi k}{a} < \sqrt{M \omega}. 
\eeq
Consistency of the EFT picture requires $M/\Lambda_{\rm sb} \ll 1$ and we see that within this regime we can neglect both types of mixing for a wide range of momenta. In particular, with some mild assumptions, we showed that in the presence of strong resonance, we can neglect the mixings between $\pi_c$ and $\chi_c$. This finding is similar in spirit to the discussion presented in the recent works \cite{Green:2014xqa,Amin:2015ftc} where those authors pointed out that it is technically natural to assume a flat field space metric in the presence of strong disorder/resonance. 

We conclude this appendix by giving the second order action for tensor perturbations and their interaction with $\pi_c$ and $\chi_c$ that we used in the main text. Using the gravitational part of the action in \eqref{EH} with \eqref{extrinsic} and noting the Ricci curvature $R^{(3)}$ on spatial hyper-surfaces,
\bea
R^{(3)} &=& \hat{g}^{ik}\partial_l\Gamma^{l}_{ik} -\hat{g}^{ik}\partial_k\Gamma^{l}_{il} + \hat{g}^{ik}\Gamma^{l}_{ik}\Gamma^{m}_{lm} -\hat{g}^{ik}\Gamma^{m}_{il}\Gamma^{l}_{km},\\
\Gamma^{k}_{ij} &=& \fr{1}{2} \hat{g}^{kl}\left(\partial_i \hat{g}_{jl}+\partial_j\hat{g}_{il}-\partial_l\hat{g}_{ij}\right),
\eea
we have the following second order action for the tensor part of the metric fluctuations
\beq
S_{\rm g} =  \fr{m_{\rm pl}^2}{8} \int \! d^4 x \:
a^3  \,  \left(\dot{h}_{ij}\dot{h}_{ij}-\fr{\partial_k h_{ij}\partial_k h_{ij}}{a^2}\right) . \;
\eeq 
On the other hand, expanding the actions \eqref{Sgpi} and \eqref{Sspi} we find the following cubic order interactions between $\pi_c$ and $\chi_c$
\beq
S_{hXX} \supset \int d^4 x~a^3 \left(\fr{c_\chi^2}{2} h_{ij} \fr{\partial_i \chi_c \partial_j \chi_c}{a^2} + \fr{c_\pi^2}{2} h_{ij} \fr{\partial_i \pi_c \partial_j \pi_c}{a^2}\right). 
\eeq

\section*{Appendix B: Relating Unitary Gauge to the Scalar Potential}
In cosmologies dominated by a scalar field, we can map the time-dependent background quantities in our Unitary gauge Lagrangian \eqref{SU} to the explicit scalar field models with a given potential $V(\phi_{_0})$. A simple example we provided in the main text was
\beq\label{PD}
V(\phi_{_0}) = m_{\rm pl}^2 (3H^2(t)+\dot{H}(t)),~~~~ -2\dot{H}m_{\rm pl}^2 =\dot{\phi}_{_0}^2
\eeq
Using $\d\phi_{_0} = \dot{\phi}_{_0} dt$ and time derivatives of expressions in \eqref{PD}, we can relate the derivatives of the potential with respect to $\phi$ to the time derivatives of the Hubble rate $H(t)$. Here, we list some of these expressions,
\bea
V'(\phi_{_0}) &=& \fr{m_{\rm pl}}{(-2\dot{H})^{1/2}}\left(6H\dot{H}+\ddot{H}\right),\\
V''(\phi_{_0}) &=& -3\dot{H} - \fr{1}{4}\left(\fr{\ddot{H}}{\dot{H}}\right)^2 -\fr{3H}{2}\left(\fr{\ddot{H}}{\dot{H}}\right)-\fr{1}{2}\partial_t\left(\fr{\ddot{H}}{\dot{H}}\right),\\
V'''(\phi_{_0}) &=& \fr{1}{(-2\dot{H}m_{\rm pl}^2)^{1/2}}\left[-\fr{H^{(4)}}{2\dot{H}}-\fr{9\ddot{H}}{2}+\fr{\ddot{H}^3}{2\dot{H}^3}-\fr{3H}{2}\partial_t\left(\fr{\ddot{H}}{\dot{H}}\right)+\fr{1}{2}\partial_t\left(\fr{\ddot{H}^2}{\dot{H}^2}\right)\right]
\eea


%

\end{document}